\documentstyle[draft%
,amscd%
,12pt]{amsart}

\newtheorem{thm}{Theorem}[section]
\newtheorem{cor}[thm]{Corollary}
\newtheorem{lem}[thm]{Lemma}

\theoremstyle{definition}
\newtheorem{defn}{Definition}[section]

\theoremstyle{remark}

\numberwithin{equation}{section}

\newcommand{\thmref}[1]{Theorem~\ref{#1}}
\newcommand{\secref}[1]{\S\ref{#1}}
\newcommand{\lemref}[1]{Lemma~\ref{#1}}
\newcommand{\bg}{\begin}
\newcommand{\dis}{\displaystyle}
\newcommand{\no}{\noindent}
\begin{document}

\title[asymptotic expansion for layer solutions]
{Asymptotic Expansion for Layer Solutions of a\\
Singularly Perturbed Reaction-Diffusion System}

\author{Xiao-Biao Lin}
\address{Department of Mathematics\\
      North Carolina State University\\
      Raleigh, North Carolina 27695--8205}
\thanks{Research partially supported by NSFgrant DMS9002803 and DMS9205535}

\email{xblin@@xblsun.math.ncsu.edu} 

\date{March 8, 1994}



\maketitle

\begin{abstract}
For a singularly perturbed system of reaction--diffusion equations,
assuming that the
0th order solutions in regular and singular regions are all stable, we
construct matched  asymptotic expansions for formal solutions to any desired
order in $\epsilon$. The formal solution shows that there is an invariant
manifold of
wave-front-like solutions that attracts other nearby solutions. With an
additional assumption on the sign of the wave speed, the wave-front-like
solutions converge slowly to stable stationary solutions on that manifold.
\end{abstract}

\include{introd}
\include{sect_b}
\include{sect2}
\include{sect3}
\include{sect4}
\include{sect5}
\include{sect6}
\include{expansion.bbl}
\end{document}


\section{Introduction}
\label{S1}

This is the first of a series of papers devoted to studying internal,
boundary and initial layers for n-dimensional systems of
reaction-diffusion equations. By a formal asymptotic method,
we derive matched expansions of layer solutions to any desired order in
$\epsilon$. We give general conditions for existence and stability of
the formal solutions. The formal expansion shows how the initial profile
quickly converges to a manifold of slow moving wave-front-like solutions.
The position of the wave front then moves slowly towards that of a stable
stationary solution. In the next
paper we will show that under the same set of conditions there is a unique
genuine solution that is near the formal series solution. For $n=1,\,2$,
these results have been obtained by other  authors. Our goal is to  generalize
their results to any finite $n$.

Consider the following reaction-diffusion equation
\begin{equation} \label{E1.1}
\epsilon u_t = \epsilon^2 u_{xx}+f(u,x,\epsilon),\quad u\in\Bbb R^n,\,
a<x<b,
\end{equation}
with Neumann boundary conditions at $x=a,b$
\begin{equation}\label{E1.1a}
u_x(a,t)=u_x(b,t)=0,
\end{equation}
 and initial condition
\begin{equation*}
u(x,0,\epsilon)=\overline u(x,\epsilon)
\end{equation*}
at $t=0$. Here $f:\Bbb R^n\times[a,b]\times\Bbb R \to \Bbb R^n$ is
$C^\infty$ with the following expansion:
$$   f(u,x,\epsilon)=\sum_{j=0}^\infty \epsilon^j f_j(u,x).
$$
Due to the presence of the small parameter $\epsilon >0$,
solutions of (\ref{E1.1}) may have internal, boundary and initial layers.
Those are the regions of $x$-$t$ space where $u_{xx}$ and/or
$u_t$ are large so that the solutions do not converge uniformly as
$\epsilon\to 0$. For the moment we ignore boundary layers, and give a
short introduction to spatially regular and internal layers. The
following is motivated by a discussion in \cite{fife76,fife76a}.

In the regular layers, a stationary solution $u(x,\epsilon)$ of (\ref{E1.1})
approaches solutions of
\begin{equation}\label{E1.2}
f_0(u,x)=0
\end{equation}
as $\epsilon\to 0$. Assume that $u=p^i(x),\,i=1,2,\,a\leq x\leq b$ are two
solutions of (\ref{E1.2}), and as $\epsilon \to 0$,
$$u(x,\epsilon)\to \begin{cases} p^1(x), & a<x<\eta,\\
                                 p^2(x), & \eta<x<b. \end{cases}
$$
There is an internal layer  at $x=\eta$. Using
stretched variables
$\xi=\frac{x-\eta}{\epsilon},\,\tau=\frac{t}{\epsilon}$,
we write (\ref{E1.1}) as (when $\epsilon=0$)
\begin{equation}\label{E1.3}
u_{\tau}=u_{\xi\xi}+f_0(u,\eta).
\end{equation}
Suppose that $\eta=\eta^0$ can be chosen such that (\ref{E1.3})
possesses a stationary  solution $q(\xi)$ that satisfies
\begin{equation}\label{E1.4}
0=u_{\xi\xi}+f_0(u,\eta)
\end{equation}
and approaches $p^1(\eta^0)$
as $\xi\to -\infty$ ($p^2(\eta^0)$ as $\xi\to\infty$), $q'(\xi)\to 0$ as
$\xi\to\pm\infty$. The functions $\{p^1(x),q(\xi),p^2(x)\}$ are the
0th order expansion of a formal solution in regular and internal
layers. The position of the stationary internal layer $\eta=\eta_0$ is
determined by the existence of a heteroclinic solution to (\ref{E1.4}).
The condition $q(\xi)\to p^i(\eta^0)$ as $\xi\to\pm\infty$
is the
0th matching condition between regular and singular layers. Higher
order matching conditions will be specified later in
this paper when higher order formal expansions are computed.

Under some general conditions, which will be
stated in \secref{S2}, it was proved \cite{lin89,lin90} that
there is an exact stationary solution $u$ to (\ref{E1.1}) near the 0th
order expansions.
Similar results were obtained in \cite{joneskopell93}.

When $\eta$ changes, generically the heteroclinic solution of (\ref{E1.4})
breaks.  However the time dependent equation (\ref{E1.3}) may have a traveling
wave solution $u(\xi,\tau)=q(\xi-V\tau,\eta)$ where $q(\xi,\eta)$ satisfies
\begin{equation}\label{E1.6}
u_{\xi\xi}+Vu_{\xi}+f_0(u,\eta)=0.
\end{equation}
Here $\eta$ serves as a parameter, and the wave speed $V=V(\eta)$ depends on
$\eta$.  The function $q(\xi,\eta)$ approaches one of
the $p^i(\eta)$, and $\frac{\partial}{\partial \xi}q$ approaches zero, as
$\xi\to\pm\infty$ due to the matching of the internal and regular layers.
The wave speed $V$ and the wave front position $\eta$ do not depend on the
stretched time $\tau$, but they depend on the slow time $t$. To see this,
let $\eta=\eta(t),\:\xi=\frac{x-\eta(t)}{\epsilon}$, and
$u(\frac{x-\eta(t)}{\epsilon},t)=u(\xi,t)$ be a solution to (\ref{E1.1}).
Then (when $\epsilon=0$)
$$
 0=u_{\xi\xi}+\eta'(t)u_\xi + f_0(u,\eta(t)).
$$
At each $t\geq 0$, we look for a heteroclinic solution of the above
connecting $p^1(\eta(t))$ to $p^2(\eta(t))$. Comparing this with
(\ref{E1.6}), we have
\begin{equation}\label{E1.7}
\frac{d\eta(t)}{dt}=V(\eta(t))
\end{equation}
We can see that (\ref{E1.7}) determines $\eta(t)$. One should not be
surprised to see that the wave speed $V$ is the same in both $x$-$t$
and $\xi$-$\tau$ coordinates, since the scaling by $\epsilon$
cancels.

Let $0<\beta<1$ be a constant. Let the width of the internal layer be
$O(\epsilon^\beta)$. The variable $x=\epsilon^\beta$ is $o(1)$ as
$\epsilon\to 0$. But in the stretched variable,
$\xi=x/\epsilon=\epsilon^{\beta-1} \to\infty$ as $\epsilon\to 0$. See
\cite{eckhaus77,eckhaus79} for discussion of such  intermediate variables.
Define a piecewise smooth function $W(x,t,\epsilon)$ by
$$
  W(x,t,\epsilon)=\bg{cases}
        p^1(x), &a < x <\eta(t)-\epsilon^\beta,\\
        p^2(x), &\eta(t)+\epsilon^\beta<x<b,\\
        q((x-\eta(t))/\epsilon),
        & \eta(t)-\epsilon^\beta<x<\eta(t)+\epsilon^\beta.
        \end{cases}
$$
At the interior of each subinterval, $W$ satisfies (\ref{E1.1}) with an
error $O(\epsilon^\beta)$. At the points $\eta(t)\pm\epsilon^\beta$, $W$ has
a jump
discontinuity of size $O(\epsilon^\beta)$. Such a function is called a
pseudo-solution to (\ref{E1.1}).
In our next paper we will show that there is an exact solution to the original
equation (\ref{E1.1}) that is near $W(x,t,\epsilon)$.
A function is said to have a wave-front-like profile
or to be a wave-front-like function if it approaches the solutions of
(\ref{E1.2}) at regular layers, but approaches heteroclinic solutions in
stretched variable at internal layers.
The function $W$ obviously has a wave-front-like profile. It follows that
the exact solution of (\ref{E1.1}) near $W$ has a wave-front-like profile.

Recall that $V(\eta_0)=0$.
The wave speed $V(\eta)$ generally changes sign when passing
$\eta=\eta^0$. The case $V<0$ if $\eta>\eta^0$ is especially interesting. It
shows that the wave-front-like solution approaches a stationary
wave-front-like solution
as time evolves. In the other case $V>0$ when $\eta>\eta^0$, the stationary
wave-front-like solution is not stable among the wave-front-like
solutions.

Suppose now the initial condition $\overline u(x,\epsilon)$ also has a
wave-front-like profile. That is, $\overline u(x,0)$ has a jump at
$x=\eta$ and is continuous on $[a,\eta)$ and $(\eta,b]$, and using the
stretched variable $\xi=\frac{x-\eta}{\epsilon}$, the limit
$\lim_{\epsilon\to 0^+} \overline u(\epsilon\xi+\eta,\epsilon)=
\overline{\overline u}(\xi,0)$ exists. We assume that as initial data
for the ODE $u_\tau=f_0(u,x)$, where $x$ serves as a parameter,
$\overline u(x,0)$ is attracted by
$p^i(x)$, $i=1\text{ for }x\in[a,\eta)$, $i=2\text{ for
}x\in(\eta,b]$, and as initial data for (\ref{E1.3}),
$\overline{\overline u}(\xi,0)$ is attracted by
$q(\xi,\eta)$ modulo a spatial shift. $q(\xi,\eta)$ is stable in the sense
of Evans, \cite{evans72a,evans72b,evans72c,evans75}. The problem of
determining when $q(\xi,\eta)$
is stable is very important, but is not the concern of this paper,
\cite{jones84,angenent87,nishiurafujii87,halesakamoto88}.

Under the above assumptions, using stretched time $\tau=t/\epsilon$,
we derive formal series solutions
in the initial layers. These solutions match with wave-front-like slow
solutions in the sense that as $\tau\to \infty$, the formal series
solutions of the
evolutionary system with the initial condition $\overline u(x,\epsilon)$
converge to the expansions in the variable $\tau$ of slow wave-front-like
solutions.

In this paper, the intermediate spaces $D_A(\theta)$ are used to study
parabolic equations. These spaces are powerful tools to treat fully nonlinear
equations.  Since our system is semi-linear, we only use some weaker
results. All the results in this paper are valid with essentially the same
proof if $D_A(\theta)$ is replaced by $D(A^\theta)$. An important feature in
this paper is to use weighted norms in function spaces. Weighted function
spaces have been used in \cite{sattinger76,sattinger77} to study the stability
of travelling waves.

The method in this paper can also handle systems with other types of
boundary conditions. We use the Neumann boundary conditions since they are
the most studied.

The outline of this paper is as follows. We introduce notations and some
basic lemmas in \S 2. In \S 3 we state a result from \cite{lin89} that
concerns the stationary wave-front-like formal series solution
(\thmref{T2.1}). The result in \S 3
is a special case of the result in \S 4. Having a separate section helps
to show what  new hypotheses are needed to study the slowly
moving wave-front-like solutions. In \S 4, we study
wave-front-like formal series solution (\thmref{T3.1}, \ref{T3.2}).
We show that these solutions have a
slowly moving front and form a manifold that depends on parameters
$\{\overline\eta_j^i\}_{j=0}^\infty,\:1\leq i\leq r-1$. Here
$\sum_0^\infty\epsilon^j \overline\eta^i_j$, is the formal series
expansion of the
initial wave front, $r-1$ is the number of internal layers. The result
partially agrees with other publications where the slow manifold is
parameterized by $\overline\eta^i_0,\;1\leq i\leq r-1$,
\cite{carrpego89,fuscohale89,halesakamoto88,alikakosbates91}.
The reason for the discrepancy is that the nonlinear function $f$ in other
papers does not depend on $x$, or $\epsilon$.
In \S 5 we study the formal series solutions in the initial layer
(\thmref{T4.1}, \ref{T4.2}). We
show that $\overline\eta^i_0$ is determined by the initial condition of
(\ref{E1.1}), and $\overline\eta^i_j,\:j\geq 1$ are determined by the
matching of initial and regular (in time) expansions. In \S 6, we prove
that the formal series solutions in adjacent layers, obtained in \S 4 and
\S 5, match with each other (\thmref{T5.1}--\ref{T5.4}). We also construct
a pseudo-solution of any
prescribed accuracy based on the matched series solutions (\thmref{T5.5}).
Most of the technical lemmas are proved in \S 7.

Internal and boundary layers in singular perturbation problems have been
an active area of current research. Various powerful methods have been
developed to treat the layer solutions
\cite{fife88,fife88a,fife89,nishiurafujii87}.
Our approach is different from the others. We
follow the pattern ``matched formal series expansions--pseudo solutions--
Newton's method'' which has been used to treat singularly perturbed ODEs,
see \cite{lin89,lin90}.
The last step uses a lemma similar to the shadowing
lemma in dynamical system theory. In our next paper we will introduce such
a shadowing lemma for parabolic systems where the pseudo-solutions have jumps
in both $x$ and $t$ directions.

\endinput


\section{Notations and basic lemmas}\label{SB}

\subsection{Notations.}

As $\epsilon \to 0$, the solution $u(x,t,\epsilon)$ of (\ref{E1.1}) may not
converge
uniformly at regions where $u_{xx}$ and $u_t$ are large.  These regions are
called singular layers with respect to space or time.  In particular, $u_t$
may
be large near $t = 0$.  That region is also called the fast (or initial)
layer
where the stretched time $\tau = t/\epsilon$ is appropriate to express
the
solutions.  $u_x$ and $u_{xx}$ may be large near some $x = \eta ^i, 0 \leq
i
\leq r$.  These regions are called boundary (internal) layers if $i = 0, r$
(or
$1 \leq i \leq r-1)$, where the stretched space variable $\xi = {x - \eta
^i\over \epsilon}$ is used.  Regions that are not singular with respect to
space or time are called regular layers.  We use $S$ or $R$ to denote
singular or regular layers.  The symbol related to space is put before that
related to time since in the dictionary order space is before time.
Thus $(SR)^i, (RR)^i, (SS)^i$ and $(RS)^i$ are used to denote the $i$th
spatially singular, temporally regular region, etc.
\vskip 0.2in
\begin{center}
\begin{tabular}{c|c|c|c|c|c|c|c|c}
t & $(SR)^0$ & $(RR)^1$ &$\ldots$ & $(RR)^i$ & $(SR)^i$ & $(RR)^{i+1}$ &
$\ldots$ &
$(SR)^r$  \\ \hline
$\tau$ & $(SS)^0$ & $(RS)^1$ &$\ldots$ &  $(RS)^i$ &
$(SS)^i$ & $(RS)^{i+1}$ & $ \ldots$ & $(SS)^r$  \\ \hline
& $\xi$ &x&$\ldots$ & x & $\xi$ & x &$\ldots$ & $\xi$
\end{tabular}
\vskip 0.2in
Figure 1
\end{center}
\vskip 0.2in
Figure 1 shows relative locations of all the possible layers.  Super-scripts
on
a solution are used to show the type of layers where the solution is
expressed
by the appropriate variables.
$$
\begin{array}{llll}
u^{RRi} (x,t,\epsilon) & = & u(x,t,\epsilon),& \mbox { for }  (x,t) \in
(RR)^i\\
u^{SRi}(\xi , t, \epsilon)  & = & u (\epsilon \xi + \eta ^i, t, \epsilon),
&\mbox { for } (x,t)\in (SR)^i\\
u^{RSi} (x,\tau , \epsilon) & = & u(x,\epsilon \tau , \epsilon), &  \mbox {
for }
(x,t) \in (RS)^i\\
u^{SSi} (\xi, \tau, \epsilon) & = & u(\epsilon \xi + \eta ^i, \epsilon
\tau , \epsilon), &  \mbox { for } (x,t) \in (SS)^i.
\end{array}
$$

Each layers is further expanded in powers of $\epsilon,\: u^{RRi}
(x,t,\epsilon)
= \displaystyle \sum ^{\infty}_{j=0} \epsilon ^iu^{RRi}_j(x,t)$, etc.

The notation $\stackrel{*}u(\tau)$ is used to denote the expansion of
$u(t)$ in the variable $\tau$. $\tilde u(\xi)$ is used to denote the
expansion of $u(x)$ in the variable $\xi$. $\overline u$ is used to denote
the initial condition for a solution $u$.

Let $C_{bu}({\Bbb R},{\Bbb R}^n), C_{bu}({\Bbb R}^+,{\Bbb R})$ and
$C_{bu}({\Bbb R}^-,{\Bbb R}^n)$ be the Banach spaces of uniformly
continuous
and bounded functions with super norms.  Let $C^m_{bu} = \{u \vert u, u',
\ldots
u^{(m)} \in C_{bu}\}$ with the norm
$$
\| u \|_{C^m_{bu}} = \sum ^m_{i=0} \| u^{(i)} \|_{C_{bu}}.
$$
Here $C^m_{bu}$ denotes $C^m_{bu}({\Bbb R},{\Bbb R}^n)$ or $C^m_{bu}({\Bbb
R}^\pm, {\Bbb R}^n)$.  We can show that $C^m_{bu}, m \geq 1$ is dense in
$C^{m-1}_{bu}$.

For a constant $\gamma\in\Bbb R$ and an integer  $j \geq 0$, define a
weight function $w(\xi)$ by
\begin{equation}\label{EB.0}
w(\xi)= (1 + |\xi|^j) e^{-\gamma \xi}.
\end{equation}

Let $E_{\Bbb R}(w)$  be the Banach space of functions with the weight $w(\xi)$.
$$
\begin{array}{c}
E_{\Bbb R}(w) = \{u:{\Bbb R}\to {\Bbb R}^n \vert u(\cdot)w^{-1}(\cdot)
 \in C_{bu}({\Bbb R},{\Bbb R}^n)\}.\\
\| u \| _{E(w)} = \sup \{|u(\xi)w^{-1}(\xi) \vert , \xi \in {\Bbb
R}\}.\\
E^m_{\Bbb R}(w) = \{u \vert u, \ldots , u^{(m)} \in E(w) \}.\\
\| u \| _{E^m(w)} = \sum ^m_{j=0}\|u^{(j)}\| _{E(w)}.
\end{array}
$$
Similar, $E^m_{{\Bbb R}^+}(w) and E^m_{{\Bbb R}^-}
(w)$ are Banach spaces of weighted functions that are defined on
${\Bbb R}^+$ and ${\Bbb
R}^-$.  We use $E^m(w)$ to denote $E^m_{\Bbb R}(w)$ or $E^m_{{\Bbb R}^{\pm}}
(w)$ if no confusion should arise.
Let
$$
\begin{array}{lll}
B^m_{{\Bbb R}^+} & = & \{ u \in C^m_{bu}({\Bbb R}^+, {\Bbb R}^n), \lim_{\xi
\to
\infty} D^\alpha_\xi u(\xi) = u^{(\alpha)}(+\infty) \mbox { exists for
$\alpha
\leq m$}\},\\
B^m_{{\Bbb R}^-} & = & \{u \in C^m_{bu}({\Bbb R}^-,{\Bbb R}^n), \lim _{\xi
\to -
\infty} D^\alpha_\xi u(\xi) = u^{(\alpha)} (-\infty) \mbox { exists for }
\alpha
\leq m\}.\\
B^m_{\Bbb R} & = & \{ u \in C^m_{bu}({\Bbb R},{\Bbb R}^n), \lim _{\xi \to
\pm
\infty} D^\alpha_\xi u(\xi) = u^{(\alpha)} (\pm \infty) \mbox { exists for
}
\alpha \leq m\}.
\end{array}
$$
One can easily verify that $u^{(\alpha)} (\pm \infty) = 0$ if $\alpha \geq
1$.

Let $w(\xi) = (1 + |\xi|^j) e^{-\gamma|\xi|},\:\gamma > 0$.
$$
\begin{array}{lll}
B^m_{{\Bbb R}^+}(w) & = \{u \in B^m_{{\Bbb R}^+},&  u(\cdot) - u(+\infty)
\in
E^m_{{\Bbb R}^+}
 (w)\}.\\
B^m_{{\Bbb R}^-} (w) & = \{u \in B^m_{{\Bbb R}^-}, &  u(\cdot) - u(-\infty)
\in
E^m_{{\Bbb R}^-}(w)\}.\\
B^m_{\Bbb R}(w)&  = \{u \in B^m_{\Bbb R}, & u(\cdot) - u(\pm \infty) \in
E^m_{{\Bbb R}^\pm} (w) \}.
\end{array}
$$

The general notation $B^m(w)$ will be used if no confusion should arise.
Let the norms of
$B^m_{{\Bbb R}^+}, B^m_{{\Bbb R} ^-}, B^m_{\Bbb R}$ be induced from
$C^m_{bu}$
respectively.  It is clear that $B^m_{{\Bbb R}^+}, B^m_{{\Bbb R}^-}$ and
$B^m_{\Bbb R}$ are Banach spaces with these norms.  Let
$$
\| u \|_{B^m_{{\Bbb R} (w)}} = \|u\|_{C^m_{bu}} + \| u - u
(+\infty)\|_{E^m_{{\Bbb
R}^+(w)}} + \|u - u(-\infty)\|_{E^m_{{\Bbb R}^-(w)}}
$$
Similarly definitions are given to $\|u\|_{B^m_{{\Bbb R}^+}}(w)$ and
$\|u\|_{B^m_{{\Bbb R}^-}} (w)$.  It can be verified that $B^m_{\Bbb
R}(w), B^m_{{\Bbb R}^-}(w)$ and $B^m_{{\Bbb R}^+}(w)$ are all Banach spaces
with the specified norms.

Let $X$ be one of the Banach spaces: $E^m_{{\Bbb R}^+}(w),\, E^m_{{\Bbb
R}^-}(w),\, B^m_{{\Bbb R}^+}(w)$ or $B^m_{{\Bbb R}^-}(w)$, with $m \geq 1$.
For
$C^1$ functions defined on $\Bbb R^+$ or $\Bbb R^-$, denote $BC = \{u \in
C^1:
u_\xi (0) = 0\}$.  $X \cap (BC)$ is a closed subspace of $X$, and a Banach
space
with the norm induced from $X$.

The following lemma is useful when working with these weighted spaces and
can be verified easily.
\bg{lem}\label{LB.0}
Let $\alpha>0,\: |\gamma|<\alpha$ be real constants, $j\geq 0$ be an
integer. Then there exists a constant $K_1$ such that
\bg{align*}
  & \int_{-\infty}^\xi e^{-\alpha(\xi-s)} (1+|s|^j)e^{-\gamma s}\,ds+
  \int_\xi^{\infty} e^{-\alpha(s-\xi)} (1+|s|^j)e^{-\gamma s}\,ds\\
  & \leq \dis \frac{K_1 e^{-\gamma\xi}}{(\alpha-|\gamma|)^{j+1}}(1+|\xi|^j).
\end{align*}
\end{lem}

\subsection{Properties of Elliptic Equations}

Assume that $f:{\Bbb R}^n \to {\Bbb R}^n$ is $C^\infty$, and $V$ is a real
constant. Equation
\begin{equation}\label{EB.1}
u_{\xi\xi} + Vu_{\xi} + f(u) = 0
\end{equation}
is equivalent to a system in ${\Bbb R}^{2n}$:
\begin{equation}\label{EB.2}
\begin{aligned}
u_\xi & =  v,\\
v_{\xi} & =  Vv - f(u).
\end{aligned}
\end{equation}
Therefore the phase space for (\ref{EB.1}) is ${\Bbb R}^{2n}$, comprising
of points $(u,u_{\xi})$.
We say $p$ is a hyperbolic equilibrium for (\ref{EB.1}) if $(p,0)$ is a
hyperbolic equilibrium for (\ref{EB.2}).  We say equation
\begin{equation}\label{EB.3}
u_{\xi\xi} + Vu_{\xi} + A(\xi)u = 0
\end{equation}
has exponential dichotomy on an interval $I\subset {\Bbb R}$ if the system
\begin{equation}\label{EB.4}
\begin{aligned} u_\xi & =  v\\
v_{\xi} & =  -Vv - A(\xi)u
\end{aligned}
\end{equation}
has an exponential dichotomy on $I$.  Here $A(\cdot):I\to {\Bbb R}^{n\times
n}$
is a continuous matrix valued function.

We say $u(\xi)$ is a heteroclinic solution of (\ref{EB.1}) if
$(u(\xi),u_\xi(\xi))$ is a heteroclinic solution for the equivalent system
(\ref{EB.2}).

\begin{lem}\label{LB.1}
Assume that $p \in {\Bbb R}^n,\, f: {\Bbb R}^n \to {\Bbb R}^n$ is $C^\infty$
and
there exists $\sigma _0 > 0$
\begin{equation}\label{EB.5}
f(p) = 0,\ R_e\sigma \{Df(p)\} \leq - \sigma _0.
\end{equation}
Then
\begin{equation}\label{EB.6}
u_{\xi\xi} + Vu_{\xi} + Df(p) u = 0
\end{equation}
has an exponential dichotomy on ${\Bbb R}$ with $n$--dimensional stable and
unstable spaces. Let $0<\alpha<\sqrt{V^2+4\sigma_0}-|V|$. Then the decay
rate on the stable (or unstable) subspace is bounded by
$Ke^{-\alpha\xi},\:\xi\geq 0$ (or $Ke^{\alpha\xi},\:\xi\leq 0$)
respectively.


Let $p^i, i = 1,2$ satisfy
(\ref{EB.5}). Let $q(\xi)$  be a solution to (\ref{EB.1}) and is defined
on ${\Bbb
R}^-$ with  $q(\xi) \to p^1$  as $ \xi \to - \infty$, and/or is defined on
${\Bbb R}^+$ with $q(\xi) \to p^2$  as $ \xi \to \infty$.  Then
\bg{equation}\label{EB.7}
u_{\xi\xi} + Vu_\xi + Df(q(\xi))u =  0
\end{equation}
has exponential dichotomies on ${\Bbb R}^-$ or  ${\Bbb R}^+$ respectively, with
$\cal {R}P_s(t)$ and $\cal {R}P_u(t)$ being $n$-dimensional subspaces in
${\Bbb
R}^{2n}$.  Here $P_u(t) + P_s(t) = I, t \in {\Bbb R}^-$ or $t \in {\Bbb
R}^+$,
are the projections to the unstable and stable subspaces. Moreover, the
decay rate $\alpha >0$ is the same as that of (\ref{EB.6}). In the case that
$q$ is a heteroclinic solution connecting $p^1$ and $p^2$, ${\cal R}P_u(0^-)
\cap {\cal R}P_s(0^+)$ is at least one dimensional, containing $(q_\xi(0),
q_{\xi\xi}(0))$.
\end{lem}

\begin{lem}\label{LB.2}
(i) Let $p \in {\Bbb R}^n$ satisfy (\ref{EB.5}), $\alpha$ be the constant
as in \lemref{LB.1}. Let $w(\xi)$ be the weight
function in (\ref{EB.0}) where $|\gamma|<\alpha$, $X = E^m_{\Bbb R} (w) $
and $g\in X$.
Then there exists a unique solution $u \in E^{m+2}_{\Bbb R}(w)$ to the
equation
$$
  u_{\xi\xi} + Vu_{\xi} + Df(p)u = g.
$$
Moreover,
$$
\|u\|_{E^{m+2}(w)} \leq C \| g\| _{E^m(w)}.
$$
(ii) Let $X$ be $E^m_{{\Bbb R}^+} (w)$ or $E^m_{{\Bbb R}^-}(w)$, and $g \in X$.
Assume that $p^i,\: i=1,2$ and $q(\xi)$ are as in \lemref{LB.1}.
Assume that $u_\xi(0) \neq 0$ for all nontrivial
bounded solutions $u$ to the equation $u_{\xi\xi} + Vu_{\xi} + Df(q)u =
0$.  Then
there exists a unique solution $u \in E^{m+2}_{{\Bbb R}^+}(w)$ or
$E^{m+2}_{{\Bbb R}^-} (w)$ to the boundary value problem
$$
\begin{array}{l}
u_{\xi\xi} + Vu_{\xi} + Df(q)u = g,\\
u_\xi (0) = \phi,
\end{array}
$$
Moreover,
$$
\|u\|_{E^{m+2}(w)} \leq C(\|g\|_{E^m(w) } + \| \phi\|_{{\Bbb R}^n}).
$$
\end{lem}

Let $p^i \in {\Bbb R}^n, i = 1,2$, satisfy (\ref{EB.5}).  Let $q(\xi)$ be a
heteroclinic solution to (\ref{EB.1}) connecting  $p^1$
to $p^2$. Let $X = E^m_{\Bbb R}(w)$ where $w(\xi)$ is as in \lemref{LB.2}.
Define $L_q:X\to X$ with $D(L_q) = E^{m+2}_{\Bbb R} (w)$ by
\begin{equation}\label{EB.8}
L_q u = u_{\xi\xi} + Vu_{\xi} + Df(q(\xi)) u.
\end{equation}

\begin{lem}\label{LB.3}
$L_q$ is a Fredholm operator with Fredholm index zero.  Assume that $\dim
Ker(L_q)=1$ then $Ker(L_q) = \mbox {span} \{q_\xi\}$ and $Range (L_q) =
\{\Psi \}^{\bot}$.  Here $\Psi$ is the unique  nontrivial bounded solution
for the adjoint equation, up to a scalar multiple,
$$
L^*_q \Psi \stackrel{def}{=} \Psi_{\xi\xi} - V\Psi_\xi + Df^\tau (q(\xi))
\Psi
= 0.
$$
$$\{\Psi\}^\bot = \{u \in X \vert \int^\infty_{-\infty} \Psi^\tau (\xi)
u(\xi)d\xi = 0\}.
$$
\end{lem}

\subsection{Properties of Parabolic Equations}

Let $A$ be a closed densely defined linear operator in a Banach space $X$.
Suppose that $A$ is sectorial and generates a $C^0$ analytic semigroup
$e^{At}$
in $X$.  For $0 \leq \theta < 1$, let $D_A(\theta)$ be the intermediate space
between $D_A$ and $X$.
$$
\begin{array}{lll}
D_A(\theta) & = & \{x \in X \vert \lim _{t \to 0} t^{1-\theta} Ae^{At} x =
0\},\\
\| x \| _{\theta} & = & \sup _{0 < t \leq 1} \vert t^{1-\theta} Ae^{At} x
\vert
_X + |x|_X,\\
D_A(\theta + 1) & = & \{x \vert x \in D_A, Ax \in D_A(\theta)\}.
\end{array}
$$
Let $D_A(1) = D_A$. Let $0\leq \beta\leq\theta \leq 1$ and $0\leq
\theta-\beta <1$.  Let $F: D_A(\theta) \to D_A(\beta)$ be a
nonlinear, Lipschitz continuous function and $x \in D_A(\theta), 0 \leq
\theta
\leq 1$.  Then there exist $t_0 > 0$ and a unique classical solution $u$
defined
on $[0,t_0]$ such that
\begin{align*}
u_t & =  Au + F(u),\\
u(0) & =  x,
\end{align*}
where
\begin{equation}\label{EB.8a}
u \in C([0,t_0]: D_A(\theta)) \cap C^1 ((0,t_0]: x) \cap C((0,t_0]: D_A).
\end{equation}
Denote the solution by $U(t)$.  Then $U:[0,t_0] \to D_A(\theta)$ is H\"older
continuous with H\"older exponent $1+\beta-\theta$.  Assume that $F \in
C^1$
with $DF$ being Lipschitz continuous.  Then $DF(U(\cdot)): [0,t_0] \to {\cal
L} (D_A(\theta): D_A(\beta))$ is H\"older continuous.  Let $g: [0,t_0] \to
D_A(\beta)$ be locally H\"older continuous, then
\begin{equation}\label{EB.8b}
\bg{aligned}
u_t & =  Au + DF(U(t)) u + g(t),\\
u(0) & =  u_0,
\end{aligned}
\end{equation}
has a unique classical solution $u$ that also satisfies (\ref{EB.8a}).
Moreover
there exists an evolution operator $T(t,s)$ such that
$$
u(t) = T(t,0) u_0 + \int^t_0T(t,s)g(s)ds.
$$
These well known facts can be found  in \cite{daprato79},
\cite{sinestrari85} and \cite{lunardi87}.

The linear equation (\ref{EB.8b}) is said to be exponentially stable if
$T(t,s)$ is defined for all $0\leq s \leq t$ and if
there exist constants $K,\,\alpha > 0,\,0\leq \theta - \beta <1$ such that
\bg{align*}
  \|T(t,s)\|_{\theta} & \leq K e^{-\alpha (t - s)}, \quad t > s;\\
  \|T(t,s)\|_{{\cal L}(D_A(\beta) : D_A(\theta))} &\leq K[1+(t-s)^{\beta-
\theta}], \quad t>s.
\end{align*}

\bg{lem}\label{LB.3a}
Assume that (\ref{EB.8b}) is exponentially stable and $g:[0,\infty)\to
D_A(\beta)$
is locally H\"older continuous and for some integer $ k \geq 0$,
$$ \|g(t)\|_{D_A(\beta)} \leq C(1+t^k), \quad t\geq 0.$$
Then for each $u_0 \in D_A(\theta)$, (\ref{EB.8b}) has a unique solution
that satisfies (\ref{EB.8a}) for all $t_0>0$ and
\[
  \|u(t)\|_{\theta} \leq C(1+t^k), \quad t\geq 0.
\]
\end{lem}

Let $w(\xi)$ be the weight functions in (\ref{EB.0}).
Let $X = E_{\Bbb R}(w)$, $E_{{\Bbb R}^+}(w)$, $E_{{\Bbb R}^-} (w)$,
$B_{\Bbb R}(w)$, $B_{{\Bbb R}^+} (w)$ or $B_{{\Bbb R}^-}(w)$.
Let $A: X \to X$ be defined as
$$
Au = u_{\xi\xi}
$$
with $D_A = E^2_{\Bbb R} (w)$ or $B^2_{\Bbb R}(w)$ if $X = E_{\Bbb R}(w)$
or
$B_{\Bbb R}(w), D_A = E^2_{{\Bbb R}^\pm} (w) \cap BC$ or $B^2_{{\Bbb
R}^\pm}
 (w) \cap BC$ if $X = E_{{\Bbb R}^\pm}
 (w) $ or $B_{{\Bbb R}^\pm} (w)$.

\begin{lem}\label{LB.4}
$A$ is a sectorial operator in $X$ with $D_A$ dense in $X$.
\end{lem}

\begin{lem}\label{LB.4a}
Let $X$ and $A:D_A\to X$ be as in \lemref{LB.4}. Then depending on the
choice of $X$, $E^1_{\Bbb R}(w)$, or $B^1_{\Bbb R}(w)$, or $E^1_{{\Bbb
R}^\pm}(w)\cap BC$, or $B^1_{{\Bbb R}^\pm}(w)\cap BC\subset D_A({1\over 2})$.
\end{lem}

Let $f: {\Bbb R}^n \to {\Bbb R}^n$ be $C^\infty$.  Assume that $f(p^1) =
f(p^2)
= 0$ and there exists $V \in {\Bbb R}$ such that equation (\ref{EB.1})
admits a
heteroclinic solution $q(\xi)$ connecting $p^1$ to $p^2$.  Let $X = E(w)$ or
$B(w)$ and $A: D_A \to X$ be defined as $Au = u_{\xi\xi}$ as in \lemref{LB.4}.
It can be verified that $f: D_A \to D_A$ and $f: X \to X$ are both $C^\infty$
Therefore, $f: D_A(\theta) \to D_A(\theta)$ for any $0 < \theta < 1$ is
also $C^\infty$.

Consider
\begin{equation}\label{EB.9}
\begin{aligned}
u_\tau & =  u_{\xi\xi} + Vu_\xi + f(u)\\
u(0) & =  \overline {u}_0,\ \overline {u}_0 \in D_A(\theta),\ 0 \leq
\theta \leq 1.
\end{aligned}
\end{equation}
For each $\overline u_0$, there exists $T > 0$ such that a
unique classical solution exists an $[0,T]$.  Also, $q(\xi)$ is a
stationary
solution to (\ref{EB.9}).  Consider the linear variational equation around
$q(\xi)$.
\begin{equation}
\label{EB.10}
\bg{aligned}
u_\tau & =  u_{\xi\xi} + Vu_\xi + Df(q(\xi))u\\
u(0) & =  \overline u_0,\ \overline u_0 \in D_A(\theta).
\end{aligned}
\end{equation}
Define $L_q u = u_{\xi\xi} + Vu_\xi + Df(q(\xi))u$.  As a perturbation to
$A$, $L_q$ is also a  sectorial
operator in $X$, cf \cite[page 80]{pazy83}.

\begin{defn}\label{DB.5}
The solution $q$ to equation (\ref{EB.9}) is said to be asymptotically
(exponentially) stable
modulo spatial shifts if there exists an open set $O \subset D_A(\theta),\,q(0)
\in
O$, such that for every $\overline u_0 \in O$, there is a constant $c \in
{\Bbb
R}$, such that $ |u(\tau) - q( \cdot + c) \vert _{D_A(\theta)} \leq
Ce^{-\gamma \tau}$.
The zero solution to equation (\ref{EB.10}) is said to be asymptotically
stable
modulo $q_\xi$ if for every $\overline u_0 \in D_A(\theta)$, there is a
constant $c \in {\Bbb R}, | u(\tau) - cq_\xi \vert _{D_A(\theta)} \leq
Ce^{-\gamma\tau}$.
\end{defn}

\begin{lem}\label{LB.6}
(Evans) (a) The stationary solution $q$ of (\ref{EB.9}) is asymptotically
stable modulo spatial shifts if and only if the zero solution of
(\ref{EB.10})
is asymptotically stable modulo $q_\xi$.

(b) The zero solution of (\ref{EB.10}) is asymptotically stable modulo
$q_\xi$
if and only if there exists $\alpha > 0$ such that
$$
\sigma \{L_q\} \cap \{Re\lambda > - \alpha \} = \{0\},
$$
and $\lambda = 0$ is a simple eigenvalue with the eigenspace spanned by
$q_\xi$.
\end{lem}

The following lemma
puts a strong restriction on the essential spectrum of $L_q$.

\begin{lem}\label{LB.7}
(a) Let $M$ be a constant $n \times n$ matrix, $\gamma\in{\Bbb R}$ be the
constant in (\ref{EB.0}) and
$Lu = u_{\xi\xi} + Vu_\xi + Mu$.  Then $L$ is a sectorial
operator on
$X = E_{\Bbb R} (w)$,  $E_{{\Bbb R}^\pm} (w)$, $ B_{\Bbb R} (w)$, or
$B_{{\Bbb R}^\pm} (w)$ 	with $D_L = E^2_{\Bbb R}(w), E^2_{{\Bbb R}^\pm} (w)
\cap BC, B^2_{\Bbb R}(w)$ or $B^2_{{\Bbb R}^\pm}(w) \cap BC$
respectively.  Assume furthermore that $Re (\sigma (M)) \leq - \sigma_0<0$
and $\gamma$ satisfy that $\eta _0 = (\gamma ^2 + 2|V\gamma|)/4 < \sigma _0$,
then
$$
Re (\sigma (L)) \leq - \sigma _0 + \eta _0
$$
(b) Let $p^i,\, i = 1,2$ satisfy (\ref{EB.5}) and $q(\xi)$ be a heteroclinic
solution connecting $p^1$ to $p^2$.  Let $X$,
$\gamma$ and $\eta_0$ be as in part (a).
Then $\sigma \{L_q\} \cap \{Re\lambda > - \sigma _0 + \eta _0\}$ consists of
only isolated eigenvalues, each is of finite algebraic multiplicity.
\end{lem}

\begin{cor}\label{CB.8}
The zero solution $u_\tau = u_{\xi\xi} + Vu_\xi + f_u(p^i) u, 1 = 1,2$, is
asymptotically stable in the space $X$.  The stationary solution $u = p^i,
i =
1,2$ for $u_\tau = u_{\xi\xi} + Vu_{\xi} + f(u)$ is asymptotically stable.
\end{cor}

\endinput


\section{Stationary solutions that have layer structures}
\label{S2}
The stationary solutions of (\ref{E1.1}) satisfy
\begin{equation}\label{E2.1}
\epsilon^2 u_{xx}+f(u,x,\epsilon)=0.
\end{equation}
The first set of assumptions are used to construct a stationary solution
to system (\ref{E1.1}) that exhibits internal and boundary layers and
to obtain the asymptotic expansion of such solution to any desired order in
$\epsilon$.

Assume that there is a partition of $[a,b]$:
$$ x^0=a<x^1<\cdots<x^r=b. $$
On each $[x^{i-1},x^i]$, a $C^\infty$ function $p^i(x),\:1\leq i \leq r$ is
defined with $f_0(p^i(x),x)=0$.

\bg{hypo}\label{H1} $Re\,\sigma\{f_{0u}(p^i(x),x)\}<0$ for $x^{i-1}\leq
x\leq x^i,\, i=1,\cdots,r$.
\end{hypo}

We introduce a stretched variable $\xi=(x-x^i)/\epsilon$ in a neighborhood
of each $x^i,\:0\leq i \leq r$. The $0$-th expansion of (\ref{E2.1}) can be
written as
\begin{equation}\label{E2.2}
u_{\xi\xi}+f_0(u,x^i)=0.
\end{equation}

Assume that a $C^\infty$ function $q^i(\xi)$ is defined for $\xi\in\Bbb R$ if
$1\leq i \leq r-1$, $\xi\in\Bbb R^+$ if $i=0$ and $\xi\in\Bbb R^-$ if $i=r$,
such that $u=q^i(\xi)$ satisfies (\ref{E2.2}). Also, $q^i(\xi)\to
p^i(x^i)$ as $\xi\to -\infty$ for $1\leq i\leq r$ and $q^i(\xi)\to
p^{i+1}(x^i)$ as $\xi\to \infty$ for $0\leq i\leq r-1$. Moreover,
$q_\xi^i(0)=0$ for $i=0,r$, cf. (\ref{E1.1a}).

The linear homogeneous equation
\begin{equation}\label{E2.3}
\phi_{\xi\xi}+f_{0u}(q^i(\xi),x^i)\phi=0,
\end{equation}
and its adjoint equation
\begin{equation}\label{E2.4}
\psi_{\xi\xi}+f_{0u}^\tau(q^i(\xi),x^i)\psi=0,
\end{equation}
are important in our study. (Here $\tau$ denotes the transpose.)

\bg{hypo}\label{H2} $q_\xi^i(\xi),\,\xi\in\Bbb R,\,1\leq i\leq r-1$ is the only
bounded solution of (\ref{E2.3}) up to  constant multiples.
\end{hypo}

Because of H2), we can show that (\ref{E2.4}) has a unique bounded
solution $\psi_i(\xi),\,\xi\in\Bbb R,\,1\leq i\leq r-1$, up to  constant
multiples. See \cite{palmer84}. Moreover, $\psi_i$ decays exponentially as
$\xi\to\pm\infty$.

\bg{hypo}\label{H3} $\dis \int_{-\infty}^\infty \psi_i^{\tau}(\xi)
f_{0x}(q^i(\xi),x^i) d\xi\neq 0,\,1\leq i\leq r-1.$
\end{hypo}

\bg{hypo}\label{H4} Let $\phi^i(\xi),\,\xi\in\Bbb R^+$ for $i=0$ and
$\xi\in\Bbb R^-$ for
$i=r$, be any nontrivial bounded solution for (\ref{E2.3}). Then
$\phi_\xi^{i}(0)\neq 0$ for $i=0,r$.
\end{hypo}

We look for the position of the internal layers $\dis x^i(\epsilon)=
\sum_{j=0}^\infty \epsilon^j x_j^i$, formal series solution $
\sum_{j=0}^\infty \epsilon^j u_j^{Ri}(x)$ to (\ref{E2.1}) in regular layer
$(x^{i-1}(\epsilon),\,x^i(\epsilon)),\,1\leq i\leq r$, and formal series
solution $\sum_{j=0}^\infty \epsilon^j u_j^{Si}(\xi),\,0\leq i\leq r$, to the
equation
\begin{equation}\label{E2.4a}
u_{\xi\xi}+f(u,x^i(\epsilon)+\epsilon\xi,\epsilon)=0,
\end{equation}
where $\xi=(x-x^i(\epsilon))/\epsilon$. Let the superscripts ``R'' and
``S'' stand for spatial regular and singular layers.
Each $u_j^{Si}(\xi)$ satisfies a growth condition
\begin{equation}\label{E2.4b}
|u_j^{Si}(\xi)|\leq C(1+|\xi|^j)
\end{equation}
as $|\xi|\to\infty$ and a boundary condition
\begin{equation}\label{E2.4c}
u_{j\xi}^{Si}(0)=0
\end{equation}
if $i=0,r$. Let the inner expansions of the outer solutions be
\begin{align*}
\sum_{j=0}^\infty\epsilon^j \tilde u_{j,1}^i(\xi)&=
 \sum_{j=0}^\infty\epsilon^j u_j^{R,i+1}(\epsilon\xi+x^i(\epsilon)),\\
\sum_{j=0}^\infty\epsilon^j \tilde u_{j,2}^i(\xi)&=
 \sum_{j=0}^\infty\epsilon^j u_j^{Ri}(\epsilon\xi+x^i(\epsilon)).
\end{align*}
We say that the solution in the singular layer $\sum\epsilon^j u_j^{Si}(\xi)$
matches solution $\sum\epsilon^j u_j^{Ri}(x)$ or $\sum\epsilon^j
u_j^{R,i+1}(x)$ if
\begin{equation}\label{E2.4d}
\begin{aligned}
|u_j^{Si}(\xi)-\tilde u_{j,1}^i(\xi)| & \leq C((1+|\xi|^j)e^{-\gamma\xi}),
\quad \xi\leq 0,\\
|u_j^{Si}(\xi)-\tilde u_{j,2}^i(\xi)| & \leq C((1+|\xi|^j)e^{-\gamma\xi}),
\quad \xi\geq 0.
\end{aligned}
\end{equation}

To construct those series solutions, we shall use the result from
\cite{lin89}. We verify that for an equivalent first order system in
$\Bbb R^{2n}$, all the conditions in \cite{lin89} are satisfied. From
\lemref{LB.1} and H1), $u=p^i(x)$ is a hyperbolic equilibrium for the equation
\begin{equation}\label{E2.5}
u_{\xi\xi} + f_0(u,x)=0,
\end{equation}
for $x^{i-1}\leq x\leq x^i$. That is, $(p^i(x),0)$ is a hyperbolic
equilibrium for the equivalent system
\begin{equation}\label{E2.6}
\begin{aligned}  u_\xi &=v,\\
v_\xi &= -f_0(u,x).
\end{aligned}
\end{equation}
Also from \lemref{LB.1}, the unstable spaces are of n-dimensional for all
$1\leq i\leq r-1$, and $x$. Thus the hypothesis H1) in \cite{lin89} is
satisfied.

On the other hand, when $x=x^i,\,1\leq i\leq r-1$, Since $q^i(\xi)\to
p^i(x^i)$ or $p^{i+1}(x^i)$ as $\xi\to -\infty$ or $ +\infty$, based on
\lemref{LB.1} again, the linearized equation

\begin{equation}\label{E2.7}
\begin{aligned} \phi_\xi &= \overline\phi,\\
         \overline\phi_\xi &= -f_{0u}(q^i(\xi),x^i)\phi,
\end{aligned}
\end{equation}
has exponential dichotomies on $\Bbb R^-$ and $\Bbb R^+$. From H2),
$(q_\xi^i(\xi), q_{\xi\xi}^i(\xi))$ is the only bounded solution of
(\ref{E2.7}) up to a constant factor. From \cite{palmer84}, there exists a
unique bounded solution $(\psi_{i\xi}(\xi),\psi_i(\xi))$, up to a constant
factor, to the adjoint equation
\begin{equation}\label{E2.8}
\begin{aligned} \psi_\xi &= f_{0u}^\tau(q_i(\xi),x^i)\overline\psi,\\
          \overline\psi_\xi &= -\psi.
\end{aligned}
\end{equation}
Cf. (\ref{E2.3}) and (\ref{E2.4}).  Denote the right hand side of
(\ref{E2.6}) by $F(u,v,x)$. Then
$$\frac{\partial F}{\partial x}(u,v,x)=\bg{pmatrix} 0\\
                                                    -f_{0x}(u,v)
                                      \end{pmatrix}.
$$
It is now clear that H3) implies \cite[H3)]{lin89}.

Finally, it is clear that H4) implies that \cite[H2)]{lin89}.
The result from \cite{lin89} yields:

\bg{thm}\label{T2.1}
Under the hypotheses H1) to H4), there exist unique formal series:
\begin{equation}\label{E2.9}
\sum_{j=0}^\infty \epsilon^j x_j^i,\quad 0\leq i\leq r,\quad x_0^i=x^i,
\,x_j^0= x_j^r=0\text{ for all } j\geq 1,
\end{equation}
\begin{equation}\label{E2.10}
\sum_{j=0}^\infty \epsilon^j u_j^{Ri}(x),\quad
u_0^{Ri}(x)=p^i(x),\quad 1\leq i\leq r,
\end{equation}
\begin{equation}\label{E2.11}
\sum_{j=0}^\infty \epsilon^j u_j^{Si}(\xi),\quad u_0^{Si}(\xi)=q^i(\xi),\quad
\text{with}\bg{cases}\xi\in\Bbb R, &1\leq i\leq r-1,\\
                     \xi\in\Bbb R^+ &i=0,\\
                     \xi\in\Bbb R^- &i=r,
           \end{cases}
\end{equation}
such that (\ref{E2.9}) is the position of the singular layer
$x^i(\epsilon)$, (\ref{E2.10}) satisfies (\ref{E2.1}), (\ref{E2.11})
satisfies (\ref{E2.4a}), (\ref{E2.4b}) and (\ref{E2.4c}) if $i=0,r$.
The functions $u_j^{Ri}$ and constants $x_j^i$ are computable by
systems of recursive linear algebraic equations. The functions $u_j^{Si}$
are computable by a system of recursive linear nonhomogeneous differential
equations. Moreover, the series solutions obtained above satisfy the
matching condition (\ref{E2.4d}).
\end{thm}

\endinput


\section{Existence of wave front like solutions} \label{S3}
\subsection{Hypotheses and lemmas}
To study time evolution solutions of (\ref{E1.1}), some more hypotheses
will be made. Let $L_{q^i}:\,C_{bu}\to C_{bu}$ be an
unbounded operator defined by
$$L_{q^i}(u)(\xi)=u_{\xi\xi}+f_{0u}(q^i(\xi),x^i)u.
$$
The domain $D(L_{q^i})=C^2_{bu}(\Bbb R,\Bbb R^n)$ for $1\leq i\leq r-1$ and
$D(L_{q^i})=C^2_{bu}\cap BC,$ for $i=0,\,r$.

\bg{hypo}\label{H5} There exists a constant $\alpha_0$ such that all the
eigenvalues of $L_{q^0}$ and $L_{q^r}$ satisfy $Re\lambda<-\alpha_0$. The
operator $L_{q^i},\,1\leq i\leq r-1$ has a simple eigenvalue $\lambda=0$
with an eigenvector $q_\xi^i(\xi)$, all the other eigenvalues satisfy
$Re\lambda<-\alpha_0$.
\end{hypo}

Hypothesis H5) ensures the stability of $q^i(\xi)$ as a solution to
$u_\tau=u_{\xi\xi}+f_0(u,x^i)$ in $C_{bu}$.

Since $\lambda=0$ is a simple eigenvalue of $L_{q^i},\,1\leq i\leq r-1$,
we have
\begin{equation}\label{E3.1}
\int_{-\infty}^\infty \psi_i^{\tau}(\xi)q_\xi^i(\xi)d\xi\neq 0,\quad 1\leq
i\leq r-1,
\end{equation}
where $\psi_i(\xi)$ is the bounded solution for (\ref{E2.4}).
The proof of (\ref{E3.1}) uses \lemref{LB.3}. If (\ref{E3.1}) were not
valid,
then $q_\xi^i$ is in the range of $L_{q^i}$. This contradicts
to the fact $\lambda=0$ is simple. Assume now

\bg{hypo}\label{H6} $\dis \int_{-\infty}^\infty
\psi_i^{\tau}(\xi)q_\xi^i(\xi)
d\xi \cdot \int_{-\infty}^\infty
\psi_i^{\tau}(\xi)f_{0x}(q^i(\xi),x^i)d\xi > 0
\quad \text{for }1\leq i\leq r-1.$
\end{hypo}

Due to H3) and (\ref{E3.1}), H6) is only a sign condition. We shall see
that H6) implies that the position of the wave front near $x^i$ moves
towards $x^i$.

\bg{lem}\label{L3.1}
There exist open intervals $O^i$ containing $x^i,\,1\leq i\leq r-1$ such
that the following holds.

\no{(i)} $p^i(x)$ and $p^{i+1}(x)$ can be extended smoothly to
$O^i$ with
$$f_0(p^j(x),x)=0,$$
$$Re\{\sigma(f_{0u}(p^j(x),x))\}<0,\quad j=i,i+1.$$
\no{(ii)} There exists a $C^\infty$ function $V^i:O^i\to \Bbb R$ such
that for each $x\in O^i$, equation
\begin{equation}\label{E3.2}
u_{\xi\xi}+V^i(x)u_\xi+f_0(u,x)=0
\end{equation}
admits a heteroclinic solution $q^i(\xi,x)$, connecting $p^i(x)$ to
$p^{i+1}(x)$, with $(u(0)-q^i(0))\bot q^i_\xi(0)$. In particular, $V^i(x^i)=0$
and $q^i(\xi,x^i)=q^i(\xi)$. Moreover, $D^k_x q^i(\cdot,x)\in C^j_{bu}$
for all $j,k\geq 0$.

\no{(iii)} The linear equation
\begin{equation}\label{E3.3}
\phi_{\xi\xi}+V^i(x)\phi_\xi+f_{0u}(q^i(\xi,x),x)\phi=0
\end{equation}
has a unique bounded solution $q^i_\xi(\xi,x)$, up to constant multiples.
And the adjoint equation
\begin{equation}\label{E3.4}
\psi_{\xi\xi}-V^i(x)\psi_\xi+f^\tau_{0u}(q^i(\xi,x),x)\psi=0
\end{equation}
has a unique bounded solution $\psi_i(\xi,x),\:|\psi_i(0,x)|=1$, up to constant
multiples. Furthermore $\psi(.,x)$ is a $C^\infty$ function of $x$ in the
space $C_{bu}$.

\no{(iv)}  In the Banach space $E_{\Bbb R}(w),\:
 L^i_x u = u _{\xi\xi} + V^i(x) u _\xi + f_{ou}(q^i(\xi,x),x) u,\:
1 \leq i \leq r-1$ has $\lambda = 0 $ as a simple eigenvalue with
eigenvector $q^i_\xi(\xi,x)$.  All the other eigenvalues of $L^i_x$ satisfy
$Re
\lambda < - \alpha _0$ for some $\alpha_0>0$.

\no{(v)} For $x\in O^i$, we have
\begin{equation}\label{E3.5}
\{\int^\infty_{-\infty} \psi^\tau_i(\xi,x)q^i_\xi(\xi,x)d\xi\} \{
\int^\infty_{-\infty}\psi^\tau_i(\xi,x)f_{0x}(q^i(\xi,x),x) d\xi\} > 0,\:
1 \leq i \leq r-1.
\end{equation}
Also $V^i(x) > 0 (=0, < 0)$ if $x < x^i(=x^i, > x^i)$. And
$$
{\partial V^i(x^i)\over \partial x} = {-\int^\infty_{-\infty} \psi ^\tau_i(\xi)
f_{0x}(q^i(\xi),x^i)d\xi\over
\int^\infty_{-\infty} \psi^\tau_i(\xi)q^i_\xi(\xi)d\xi}<0.
$$
\end{lem}

The proof of \lemref{L3.1} will be given in \S 6.

Consider the nonhomogeneous equation, $1 \leq i \leq r - 1$,:
\begin{equation}\label{E3.6}
u_{\xi\xi} + V^i(x) u_\xi + f_{0u}(q^i(\xi,x), x)u + V_1 q^i_\xi (\xi,x) =
g(\xi).
\end{equation}
Here $V^i(x)$ and $q^i(\xi,x)$ are the functions in (iii),
$V_1 \in {\Bbb R}$ is a parameter.
Suppose that $g \in E^m_{\Bbb R} (1 + |\xi|^k), m, k \geq 0$.

\begin{lem}\label{L3.2}

Let $1 \leq i \leq r - 1$.  There exists a unique $C^\infty$ function
$V^i_*:
O^i \times E^m_{\Bbb R}(1 + |\xi|^k) \to {\Bbb R}$ such that if $V_1
=
V^i_*(x,g)$, then there is a unique solution $u(\xi,x,g)$ of (\ref{E3.6})
with
$u \in E^{m+2}_{\Bbb R} (1 + |\xi|^k)$ and $u(0)\perp q^i_\xi(0 ,x) $.
Moreover, $u (\cdot , x,g)$ is $C^\infty$ in $(x,g)$ with respect to the
indicated norms.

\end{lem}

\subsection{Formal power series solutions in ${\bold (RR)^i}$}
Let the position of the $i$-th internal layer be
\begin{equation}\label{E3.7}
\eta ^i(t,\epsilon)  =  \sum ^\infty_{j=0} \epsilon ^j\eta ^i_j(t),\quad
 1 \leq i \leq r-1.
\end{equation}
For convenience, let $\eta ^0(t,\epsilon) = a,\; \eta ^r(t,\epsilon) = b$.

Assume that each $\eta ^i_0(t) \in O^i$.  In the interval
$(\eta^{i-1}_0(t),
\eta ^i_0(t)), \ 1 \leq i \leq r$, we seek formal series solution
$$
u^{RRi} (x,t,\epsilon) = \sum ^\infty_{j=0}\epsilon ^ju^{RRi}_j(x)
$$
that satisfies (\ref{E1.1}).  Since $u_t = 0$, expanding in powers of
$\epsilon$, we have (drop  the super--scripts):
\begin{align}
0 & =  f_0(u_0(x),x)\label{E3.8}\\
0 & =  f_{ou}(u_0(x),x)u_1 +
f_1(u_0(x),x)\tag{\ref{E3.8}${}_1$}\\
&  \ldots\notag
\end{align}
\begin{equation}
0 = f_{ou}(u_0(x),x)u_k + u_{k-2,xx} + \sum
_{\alpha,\delta} C_{\alpha \delta}
D^{|\alpha|}_uf_\delta(u_0(x),x)u^\alpha.\tag{\ref{E3.8}${}_k$}
\end{equation}
Here $\alpha = (\alpha _1,\ldots,\alpha_{k-1})$ is a multi-index, $u^\alpha =
u^{\alpha_1}_1 \ldots u^{\alpha_{k-1}}_{k-1},\: \displaystyle \sum
^{k-1}_{j=1} j \alpha _j + \delta =k,\: C_{\alpha \delta}$ is a constant.

Let the solution of (\ref{E3.8}) be $u_0(x) = p^i(x), 1 \leq i \leq r$.
{}From H1) and \lemref{L3.1} (i), $f_{ou}(u_0(x),x)$ is nonsingular.  Thus,
$u^{RRi}_j,\, j \geq 1$ can be solved successively from system (\ref{E3.8})--
(\ref{E3.8}${}_k$), $k\geq 1$.

\bg{thm}\label{T3.1}
Assume H1) and $\eta^i_0(t)\in O^i$ for $1\leq i \leq r-1$. Then there exists a
unique formal series solution
$u^{RRi} (x,t,\epsilon) = \sum ^\infty_{j=0}\epsilon ^ju^{RRi}_j(x)$ that
formally satisfies
$$
  0=\epsilon^2 u_{xx}+f(u,x,\epsilon),\quad \eta_0^{i-1}(t)<x<\eta_0^i(t).
$$
The solutions are independent of $t$, With $u^i_0=p^i(x)$, they  can be
obtained from system (\ref{E3.8})--(\ref{E3.8}${}_k$), $k\geq 1$ recursively.
\end{thm}

\subsection {Formal solutions in ${\bold (SR)^i}$} First consider the internal
layers, $1\leq i\leq r-1$. Let the position of the interval layer,
at $t = 0$, be $\eta ^i(0,\epsilon)=
\displaystyle \sum ^\infty_{j=0} \epsilon ^j\overline \eta ^i_j$, i.e.,
$\eta
^i_j(0) = \overline \eta ^i_j$. Assume in this section that
$\{\overline\eta_j^i\}_0^\infty$ is given. The problem of determining
$\{\overline\eta_j^i\}_0^\infty$ will be
discussed in \S 5. Assume that
\begin{equation}\label{E3.9}
\overline \eta ^i_0 \in O^i,\ 1 \leq i \leq r - 1\quad \mbox { and }\quad
\overline \eta ^{i-1}_0 < \overline \eta ^i_0,\ 2 \leq i\leq r - 1.
\end{equation}
Let $\xi = [x - \eta ^i(t,\epsilon)]/\epsilon$.  We seek the layer
position
$\eta ^i(t,\epsilon)$ and the formal solution $u^{SRi}(\xi,t,\epsilon)$
near
the singular layer at $x = \eta ^i(t,\epsilon)$.  Since $u(x,t,\epsilon) =
u^{SRi}((x - \eta ^i(t,\epsilon))/\epsilon, t, \epsilon)$, from
(\ref{E1.1}),
$u^{SRi}(\xi,t,\epsilon)$ satisfies (drop the super-indices):
\begin{equation}\label{E3.9a}
\epsilon u_t = u_{\xi\xi} + D_t\eta (t,\epsilon)u_\xi + f(u,\epsilon\xi + \eta
(t,\epsilon),\epsilon).
\end{equation}
Let $u(\xi,t,\epsilon) = \displaystyle \sum ^\infty_{j=0} \epsilon
^ju_j(\xi,t),\: \eta (t,\epsilon) = \displaystyle \sum ^\infty_{j=0} \epsilon
^j\eta _j(t)$.  Expanding in powers of $\epsilon$, we have ($\eta '(t)$
denotes
${d\over dt} \eta (t)$):
\begin{align}
0 & =  \eta '_0(t) u_{0\xi} + u_{0\xi\xi} + f_0(u_0,\eta
_0(t)),\label{E3.10}\\
u_{0t} & =  \eta '_1(t) u_{0\xi} + f_{0x}(u_0,\eta _0(t)) \eta
_1(t)\nonumber\\
& +  u_{1\xi\xi} + \eta '_0(t) u_{1\xi} + f_{0u}(u_0,\eta
_0(t))u_1\tag{\ref{E3.10}${}_1$}\\
& +  \{f_{0x}(u_0,\eta _0(t)) \xi + f_1(u_0,\eta _0(t))\},\nonumber
\end{align}
$$
\ldots
$$
\begin{align}
u_{k-1,t} & =  \eta '_k(t) u_{0\xi} + f_{0x}(u_0,\eta _0(t))\eta
_k(t)\nonumber\\
& +  u_{k\xi\xi}  +  \eta '_0(t)u_{k\xi} +
f_{0u}(u_0,\eta_0(t))u_k\tag{\ref{E3.10}${}_k$}
\\
& +  \sum ^{k-1}_{j=1}\eta '_j(t)u_{k-j,\xi} + \sum  C_{\alpha \beta
\gamma
\delta} D^{|\alpha|}_x f_\delta (u_0,\eta _0(t))u^\alpha \eta ^\beta \xi
^\gamma.\nonumber
\end{align}
Here $\alpha = (\alpha _1, \ldots , \alpha _{k-1}),\: \beta = (\beta _1,
\ldots ,
\beta _{k-1})$ are multi-indices, $\delta$ and $\gamma$ are nonnegative
integers, $u^\alpha = u_1^{\alpha _1} \ldots u^{\alpha _{k-1}}_{k-1},\: \eta
^\beta = \eta ^{\beta _1}_1 \ldots \eta ^{\beta _{k-1}}_{k-1},\: \delta +
\gamma
+ \displaystyle \sum ^{k-1}_{j=1} (\alpha _j + \beta _j) \cdot j = k,\:
C_{\alpha \beta \gamma \delta}$ is a constant.

{}From \lemref{L3.1}, (ii), there exists a unique heteroclinic solution
\begin{equation}\label{E3.11a}
u_0(\xi,t) = q^i(\xi,\eta _0(t))
\end{equation}
to (\ref{E3.10}) connecting $p^i(\eta ^i_0(t))$ to $p^{i+1}(\eta ^i_0(t))$
with
$(u_0(0,t)-q^i(0)) \bot  q^i_\xi(0)$, provided that
\begin{equation}\label{E3.11}
\eta '_0(t) = V^i(\eta _0(t)).
\end{equation}
With the initial condition $\eta ^i_0(t) = \overline \eta ^i_0$,
(\ref{E3.11})
uniquely determines   $\eta ^i_0(t), t \geq 0$.  From \lemref{L3.1}, (v), $x =
x^i$ is the stable equilibrium of (\ref{E3.11})  in $O^i$.  Since
$\overline
\eta ^i_0 \in O^i$, we have $\eta ^i_0(t) \in O^i$ for all $t \geq 0$ and
approaches $x^i$ as $t \to \infty$.  From (\ref{E3.11a}) and the last
assertion of \lemref{L3.1}, (ii), and (\ref{E3.11a}), we see that $D^j_t u_0
\in C^m_{bu}$  for all $j,\, m \geq 0$.

We compute the sequences $\{u_j\}^\infty_{j=0}$ and $\{\eta
_j\}^\infty_{j=0}$ by induction.  Assume that $u_j$ and $\eta _j,\,
0 \leq j \leq k -1$, have been obtained, $u_j$ is written as $u_j(\xi,t) =
U_j(\xi,x,x_1,\ldots,x_j)$ where $x = \eta _0(t),\: x_1 =\eta_1(t), \ldots ,
x_{k-1} = \eta _{k-1}(t)$, and $D^\alpha _yU_j \in E^m(1 + |\xi|^j)$ where
$y =
(x, x_1, \ldots , x_j),\: \alpha = (\alpha _0, \alpha _1, \ldots , \alpha
_{j-1}),\: m \geq 0$ is an arbitrary integer. Also assume that $\eta '_j(t) =
V^i_j(x,x_1,\ldots,x_j)$ and $\eta _j(t) \to x^i_j$ as $t \to \infty$.  Let
$\eta _k(t) = x_k$ and $\eta '_k(t) = V_k$.  We can write $u_{k-1,t} =
\displaystyle \sum ^{k-1}_{\ell =  0} {\partial U_{k-1} \over \partial
x_\ell}\cdot V^i_\ell(x, \ldots , x_\ell)$ where $x_0  = x$ and $V^i_0(x) =
V^i(x)$.  Equation (\ref{E3.10}${}_k$) can be written as
\begin{equation}\label{E3.12}
\bg{aligned}
u_{k\xi\xi} & + V^i(x) u_{k\xi} + f_{ou} (q^i(\xi,x),x)u_k +
V_kq^i_\xi(\xi,x)\\
& = h_k(\xi,u_0,u_1,\ldots,u_{k-1},x,x_1,\ldots,x_k).
\end{aligned}
\end{equation}
{}From the induction assumptions, we can verify that $D^\alpha _yh_k \in
E^m(1 +
|\xi|^k)$ for all $m \geq 0$, where $\alpha  = (\alpha _0, \ldots , \alpha
_k)$
and $y = (x, \ldots , x_k)$.  Therefore by \lemref{L3.2}, there exists a
unique
$C^\infty$ function $V^i_k: O^i \times {\Bbb R}^k \to {\Bbb R}$ such that
if
$V_k = V^i_k(x,x_1,\ldots,x_k)$, then (\ref{E3.12}) has a unique
solution $U_k(\xi,x,x_1, \ldots ,x_k)$, $ U_k(0,x,\ldots,x_k) \bot
q^i_\xi(0,x)$ and
$$
D^\alpha _yU_k \in E^m(1 + |\xi|^k)
$$
for all $m \geq 0$.  With the initial condition $x_k(0) = \overline \eta
^i_k$,
equation
\begin{equation}\label{E3.13}
x'_k = V^i_k(x,x_1,\ldots,x_k)
\end{equation}
has a unique solution $x_k= \eta _k(t)$.  Let
\begin{equation}\label{E3.13a}
u_k(\xi,t) = U_k(\xi,\eta _0(t), \ldots , \eta _k(t)).
\end{equation}
This is clearly a solution to (\ref{E3.10}${}_k$) and satisfies
\bg{equation}\label{E3.19}
|D^\ell_t u^i_j |\in E^m(1+|\xi |^j),\quad \ell,\,m\geq 0.
\end{equation}
Since $h_k$ is linear in
$x_k$
with ${\partial h_k \over \partial x_k} = - f_{0x}(u_0,x)$, and $\{h_k -
V_kq^i_\xi(\xi,x)\} \in \{\psi _i(\cdot , x)\}^\bot$, see  the proof of
\lemref{L3.2}, thus $V^i_k$ is a linear function of $x_k$ and
$$
{\partial V^i_k \over \partial x_k} = - \left \{ \int ^\infty _{-\infty}
\psi
^\tau _i(\xi,x)q^i_\xi(\xi,x)d\xi\right \} ^{-1} \cdot
\left \{ \int^\infty_{-\infty}
\psi^\tau_i(\xi,x)f_{0x}(q^i(\xi,x),x)d\xi\right
\} < 0.
$$
Recall that $\eta _j(t) \to x^i_j, \ 0 \leq j \leq k - 1$ as $t \to
\infty$.
Also, when $x_j = x^i_j, 0 \leq j \leq k -1,\: x_k = x^i_k$ is a stable
equilibrium solution to (\ref{E3.13}) and $\eta _k(t) \to x^i_k$ as $t \to
\infty$.  See \thmref{T2.1}  for $\left \{ x^i_j\right \} ^k_{j=0}$.  Finally
$u_k(\xi,t) \to U_k(\xi, x^i_0, \ldots , x^i_k)$ as $t \to \infty$, the
latter
is $u^{Si}_k(\xi)$  in \thmref{T2.1}.

Next, consider the boundary layers, $i=0,\,r$.
At the boundary layer $(SR)^0$ near $x = a$, let $x = a + \epsilon \xi,\:
u(x,t,\epsilon) = u^{SR0}(\xi,t,\epsilon) = \displaystyle \sum
^\infty_{j=0}
\epsilon ^ju^{SR0}_j(\xi,t)$.  The system for $\{u^{SR0}_j\}$ is simpler
than
(\ref {E3.10})--(\ref{E3.10}${}_k$) since the layer position does not move.
After dropping the super--indices, we have
\begin{equation}
0 = u_{0\xi\xi} + f_0(u_0,a),\label{E3.14}
\end{equation}
\begin{equation}
u_{0t} = u_{1\xi\xi} + f_{0u} (u_0,a)u_1 + f_{0x}(u_0,a) \xi + f_1(u_0,a),
\tag{\ref{E3.14}${}_1$}
\end{equation}
\begin{equation}\tag{\ref{E3.14}${}_k$}
u_{k-1,t} = u_{k\xi\xi} + f_{0u}(u_0,a)u_k + \sum C_{\alpha \gamma \delta}
D^{|\alpha|}_uD^\gamma _xf_\delta(u_0,a)u^\alpha \xi ^\gamma.
\end{equation}
Here $\alpha = (\alpha _1, \ldots , \alpha _{k-1}),\, u^\alpha  = u^{\alpha
_1}_1
\ldots u^{\alpha _{k-1}}_{k-1},\, \gamma \geq 0 ,\, \delta + \gamma +
\displaystyle
\sum ^{k-1}_{j=1} j \alpha _j = k$ and $C_{\alpha \gamma \delta}$ is a
constant.  The boundary conditions
\begin{equation}\label{E3.15}
u_{j\xi} (0) = 0
\end{equation}
are imposed on $\{u_j\}^\infty_{j=0}$.

Since $q^0(\xi)$ satisfies (\ref{E3.14}) and (\ref{E3.15}), set
$u_0(\xi,t) =
q^0(\xi)$.  $u_0 \in C^m_{bu} ({\Bbb R}^+,{\Bbb R}^n)$ for all $m \geq 0$.
Observe that the right hand side of (\ref{E3.14}${}_j$) does not depend on $t$.
Thus, $u_{jt} = 0$ for all $j \geq 0$.  After rewriting (\ref{E3.14}${}_j$) and
(\ref{E3.15}) to a first order system in  ${\Bbb R}^{2n}$, we find they
correspond
to (5.10j) in \cite[\S5]{lin89}.  We look for a solution satisfying
\begin{equation}\label{E3.16}
u_j \in E^m_{{\Bbb R}^+} (1 + |\xi|^j).
\end{equation}
This condition correspond to (5.11j) in \cite{lin89}.  Moreover
Hypothesis
H4) implies assumption (H2) in \cite{lin89}.  From the results of \cite{lin89},
we
conclude that there exists $\{u^{SR0}_j\}^\infty_{j=0}$ satisfying
(\ref{E3.14})--(\ref{E3.14}${}_j$), $j\geq 1$ and (\ref{E3.15}) and the
growth condition (\ref{E3.16}).  In
fact $u^{SR0}_j(\xi,t) = u^{Si}_j(\xi)$ as in \thmref{T2.1} of this paper.
Similar arguments also apply to $(SR)^r$.

\bg{thm}\label{T3.2}
Assume that H1)--H6) and $\sum_{j=0}^\infty \epsilon^j
\overline\eta_j^i$ is given that satisfies (\ref{E3.9}), $1\leq i\leq
r-1$, and $\sum \epsilon^j\overline\eta^0_j = a$, $\sum
\epsilon^j\overline\eta^r_j = b$. Then there exist unique formal series
\bg{equation}\label{E3.17}
\eta^i(t,\epsilon)=\sum\epsilon^j\eta^i_j(t),\: 0\leq i\leq r,\quad
\eta^i_j(0)=\overline\eta^i_j,
\end{equation}
\bg{equation}\label{E3.18}
u^{SRi}(\xi,t,\epsilon)=\sum\epsilon^j u_j^{SRi}(\xi,t),\:0\leq i\leq
r,\quad u_0^{SRi}(\xi,t)=q^i(\xi,\eta_0^i(t)),
\end{equation}
with $u^{SRi}_j$ defined for $\xi\in {\Bbb R}$ if $1\leq i\leq r-1$,
$\xi\in {\Bbb R}^+$ if $i=0$, $\xi\in {\Bbb R}^-$ if $i=r$ such that the
followings are satisfied:
(\ref{E3.17}) and (\ref{E3.18}) formally satisfy (\ref{E3.9a}).
$u^{SRi}_j(\xi,t)$ satisfies (\ref{E3.19}),
and the boundary condition (\ref{E3.15}) if $i=0,r$.

The series $\sum\epsilon^j u^{SRi}_j$ is computable recursively from
(\ref{E3.10})--(\ref{E3.10}${}_k$), $k\geq 1$, and
$\sum\epsilon^j\eta^i_j(t)$ is computable recursively form (\ref{E3.11})
if $j=0$ or (\ref{E3.13}) if $j\geq 1$.

Moreover, $\eta^i_j(t)\to x^i_j$ as $t\to\infty$, where $\sum\epsilon^j
x^i_j$ is the stationary front as in \thmref{T2.1}. Also $u^{SRi}_j
(\xi,t) \to  u^{Si}_j(\xi)$ as $t\to \infty$, where $\sum\epsilon^j u^{Si}_j
(\xi)$ is the formal series solution in the stationary singular layer as in
\thmref{T2.1}.
\end{thm}

\endinput


\section{Solutions in the initial layer}
\label{S4}

In the initial layer near $t = 0$, we use the stretched time $\tau =
t/\epsilon$.  (\ref{E1.1}) is now
\begin{equation}\label{E4.1}
u_\tau = \epsilon ^2u_{xx} + f(u,x,\epsilon).
\end{equation}

\subsection {Assumptions on the initial conditions}

We assume that the initial data $\overline u(x,\epsilon)$ has a layer
structure described as follows.  There is a partition of the interval $[a,b]$
$$
a = \overline \eta ^0 < \overline \eta ^1 < \ldots < \overline \eta ^r = b
$$
with $\overline \eta ^i \in O^i,\, 1 \leq i \leq r - 1$.  At each $\overline
\eta ^i,\, 1 \leq i\leq r -1$, using the stretched variable $\xi = (x
-\overline
\eta ^i)/\epsilon$, we have $ \overline u(x,\epsilon) =
\overline u(\overline\eta ^i +
\epsilon \xi , \epsilon) = \overline{\overline u}^{Si}(\xi,\epsilon),\, \xi \in
{\Bbb R}$ if $1 \leq i \leq r - 1, \xi \in {\Bbb R}^+$ if $i = 0,\: \xi
\in {\Bbb R}^-$ if $i = r$. Assume that

\bg{hypo}\label{H7}
$$
\overline u(x,\epsilon) = \sum ^\infty_{j=0} \epsilon ^j\overline
u^{Ri}_j(x),\quad \overline \eta ^{i-1}<x<\overline \eta ^i,\ 1 \leq i\leq r
$$
$$
\overline {\overline u} (\xi,\epsilon) = \sum ^\infty_{j=0} \epsilon ^j
\overline {\overline u}^{Si}_j(\xi),\quad 0 \leq i \leq r.
$$
$$
\overline {\overline u}^{Si}_{j\xi} (0) = 0,\quad j \geq 0,\:i=0,r.
$$
The functions $\overline u^{Ri}_j$ and $\overline {\overline
u}^{Si}_j$ are $C^\infty$.  $\overline u^{Ri}_j(x)$ has a $C^\infty$
extension
to $[\overline \eta ^{i-1}, \overline \eta ^i]$.
\end{hypo}
Let the inner expansion of the outer power series be
$$
\begin{array}{lll}
\sum ^\infty_{j=0} \epsilon ^j\tilde{\tilde u}^{Ri}_{j,1}(\xi) & = & \sum
^\infty_{j=0} \epsilon ^j \overline u^{Ri}_j(\overline \eta ^i + \epsilon
\xi), \quad 1 \leq i \leq r,\\
\sum ^\infty _{j=0} \epsilon ^j\tilde {\tilde u}^{Ri}_{j,2} (\xi) & = &
\sum
^\infty _{j=0} \epsilon ^j\overline u^{R(i+1)}_j (\overline \eta ^i +
\epsilon
\xi), \quad 0 \leq i \leq r-1.
\end{array}
$$
Here, $\tilde {\tilde u}^{Ri}_{j,\nu}(\xi), \nu = 1,2$, is a polynomial of
degree $j$.

\bg{hypo}\label{H8}  Each $\overline {\overline u}^{Si}_j \in E^m(1 +
|\xi|^j)$,
for all $m \geq 0$, and the matching conditions with outer expansions are
satisfied:
$$\begin{array}{ll}
\overline {\overline u}^{Si}_j(\xi) - \tilde{\tilde u}^{Ri}_{j,1} (\xi)
\in
E^m_{{\Bbb R}^-} ((1 + |\xi|^j) e^{-\gamma|\xi|}),\\
\overline {\overline u}^{Si}_j
(\xi) - \tilde {\tilde u}^{Ri}_{j,2}(\xi) \in E^m_{{\Bbb R}^+} ((1 +
\xi^j)
e^{-\gamma \xi}),
\end{array}
$$
where $\gamma > 0$ is a constant.
\end{hypo}

General discussion of matching conditions can be found in
\cite{eckhaus77,eckhaus79}.

In the regular region, the $0$-th order equation of (\ref{E4.1}) is
\begin{equation}\label{E4.2}
u_\tau = f_0(u,x),\quad x \in [\overline \eta
^{i-1},\overline \eta ^i].
\end{equation}
Let $V^i(x)$ be the function in \lemref{L3.1}, (ii) so that the following
equation
\begin{equation}\label{E4.3}
u_\tau = u_{\xi\xi} + V^i(\overline \eta ^i)u_\xi + f_0(u,\overline \eta
^i)
\end{equation}
has a stationary solution (heteroclinic solution, $u_\tau = 0)\
q^i(\xi,\overline \eta ^i)$ connecting $p^i(\overline \eta ^i)$ to
$p^{i+1}(\overline \eta ^i),\, 1 \leq i \leq r - 1$.  For $i=0,r$,
let $q^i(\xi,\overline \eta^i) = q^i(\xi),\: V^i(\overline \eta ^i) = 0$.
Recall that $q^0(\xi),\,
\xi \geq 0$ and $q^r(\xi),\, \xi \leq 0$ are stable stationary solutions of
(\ref{E4.3}) satisfying boundary conditions at $x = a$ or $x = b$,
approaching
$p^1(a)$ as $\xi \to + \infty$ or $p^r(b)$ as $\xi \to - \infty$
respectively.

According to \lemref{LB.4}, $D^2_\xi$ is a sectorial operator in $X =
{\Bbb B}_{\Bbb R}
(w)$, so is its perturbation $Au = u_{\xi\xi} + V^i(\overline \eta
^i)u_\xi,\, 1
\leq i \leq r - 1$.  When $i = 0,r$, $A$ is sectorial in $X = B_{{\Bbb
R}^\pm}
(w)$ with $D_A = B^2_{{\Bbb R}^\pm}(w) \cap (BC)$.  Observe that $u
\to f_0(u,\overline \eta ^i)$ maps $D_A$ to itself.  Let $u(0) =
u_0 \in D_A$.  Local existence for solutions of (\ref{E4.3}) in $B_{\Bbb
R}(w),\,
1 \leq i \leq r-1$ or $B_{{\Bbb R}^\pm}(w),\, i = 0,r$ has been established.
In
particular, $u \in C^1 ([0,t_0]: X) \cap C([0,t_0]: D_A)$.  See
\cite{daprato79}

\bg{hypo}\label{H9} (i) For each $x \in [\overline \eta ^{i-1},\overline \eta
^i]$, the equilibrium $p^i(x)$ of (\ref{E4.2}) attracts $\overline
u^i_0(x)$.\newline
(ii) The stationary solution $q^i(\xi)$ of (\ref{E4.3}) attracts
$\overline {\overline u}^{Si}_0(\xi)$ in the space $B^2_{{\Bbb R}^+}(w)$
or
$B^2_{{\Bbb R}^-}(w)$ if $i = 0,r$; the stationary solution
$q^i(\xi,\overline
\eta ^i)$ of (\ref{E4.3}) attracts $\overline {\overline u}^{Si}_0(\xi)$
in
the space $B^2_{\Bbb R}(w)$ modulo a spatial shift if $1 \leq i \leq r -
1$.
\end{hypo}

H\ref{H9}), (i) is a reasonable assumption since  H1) and
\lemref{L3.1}, (i)  imply that $p^i(x)$ is a stable solution for (\ref{E4.2}).
H9), (ii) is also a reasonable assumption since from H8), we have
$\overline {\overline u}_0^{Si}\in B^2(w)$ and we can prove
 the following lemma.

\begin{lem}\label{L4.1}
$q^0(\xi,\overline \eta ^0)$ and $q^r(\xi,\overline \eta ^r)$ are
asymptotically stable stationary solutions in $B^2_{{\Bbb R}^+}
(w)$ and $B^2_{{\Bbb R}^-}(w)$ respectively.  $q^i(\xi,\overline \eta ^i),
\,1 \leq i \leq r -1$ are asymptotically stable in $B^2_{\Bbb R}(w)$
modulo spatial shifts.
\end{lem}

For $x \in O^i$, define
$$
L_{q^i}u = u_{\xi\xi} + V^i(x) u_\xi + f_{0u}(q^i(\xi,x),x)u,\quad 0 \leq
i \leq r.
$$
See \lemref{L3.1} (ii) for $V^i(x)$ and
$q^i(\xi,x)$.  $L_{q^i}$ is a closed linear operator if $X = E_{\Bbb R}(w),\,
1 \leq i \leq r - 1$ with $D(L_{q^i}) = E^2_{\Bbb R}(w)$;
or if $X  = E_{{\Bbb R}^\pm} (w),\, i = 0$ or $r$, with $D(L_{q^i}) =
E^2_{{\Bbb R}^\pm} (w) \cap (BC)$.  Consider
\begin{equation}\label{E4.3a}
u_\tau  = L_{q^i}u,\quad   0 \leq i \leq r.
\end{equation}
Also the boundary condition $u_\xi (0) = 0$ is imposed if $i = 0,r$.

\begin{lem}\label{L4.2}
Equation (\ref{E4.3a}) is asymptotically stable in $E^2_{{\Bbb R}^\pm} (w)$
if
$i = 0,r$.  It is asymptotically stable modulo $q^i_\xi(\cdot,x)$ if $1
\leq i
\leq r - 1$.
\end{lem}

\subsection{Formal power series solutions in ${\bold (RS)^i}$}
Let $u^{RSi}(x,\tau,\epsilon) = \displaystyle \sum ^\infty_{j=0}\epsilon
^ju^{RSi}_j(x,\tau)$.  From (\ref{E4.1}), expanding in powers of
$\epsilon$ and dropping the super indices,
we have for $x \in [\overline \eta ^{i-1},\overline \eta ^i]$,
\begin{align}
u_{0\tau} & =  f_0(u_0,x),\label{E4.4}\\
u_{1\tau} & =  f_{0u}(u_0,x)u_1 + f_1(u_0,x),\tag{\ref{E4.4}${}_1$}
\\
u_{2\tau} & =   u_{0xx} + f_{0u}(u_0,x)u_2 + f_{0uu} \cdot u^2_1/2 +
f_{1u}
\cdot u_1 + f_2, \tag{\ref{E4.4}${}_2$} \\
& \ldots \nonumber \\
u_{k\tau} & =  u_{k-2,xx} + f_{0u}(u_0,x)u_{k} + \sum  C_{\alpha \delta}
D^{|\alpha|}_uf_\delta \cdot u^\alpha.
\tag{\ref{E4.4}${}_k$}
\end{align}
Here $\delta \geq 0,\, \alpha = (\alpha _1, \ldots , \alpha _{k-1}),\,
u^\alpha =
u^{\alpha _1}_1 \ldots u^{\alpha _{k-1}}_{k-1},\, \delta + \sum j\alpha _j =
k,\, C_{\alpha \delta}$ is a constant.

With $x$ as a parameter, (\ref{E4.4}) - (\ref{E4.4}${}_k$), $k\geq 1$,
are to be solved recursively with the initial data
\begin{equation}\label{E4.5}
u_j(x,0) = \overline u_j(x).
\end{equation}

Since $p^i(x)$ attracts $\overline u_0(x)$, cf. H9), (i), the solution
$u^i_0(x,\tau)$ of (\ref{E4.4}) approaches $p^i(x)$ exponentially a
$\tau \to
\infty$.  From a standard perturbation theory, the linear variational
equation
$$
u_\tau = f_{0u}(u_0,x)u
$$
is exponentially stable.  We now proceed by induction.  Assume that $u_j, 0
\leq j \leq k - 1$ have been solved with $|D^\alpha
_xu_j(x,\tau)|_{C^m([0,\infty))} \leq C_{j\alpha m}$, for all $\alpha \geq
0, m
\geq 0$.  Rewrite (\ref{E4.4}${}_k$) as
\begin{equation}\label{E4.6}
u_{k\tau} = f_{0u}(u_0,x)u_k + h_k(u_0,u_1, \ldots , u_{k-1})
\end{equation}
It is easy to see that $|D^\alpha _xh_k|_{C^m([0,\infty))} \leq \infty $
for all
$\alpha\geq 0, m \geq 0$.  Equation (\ref{E4.6}) with initial condition
(\ref{E4.5}) then has a unique solution $u_k$ that satisfies
\begin{equation}\label{E4.7}
|D^\alpha _x u_k(x,\tau)|_{C^m([0,\infty))} \leq C,
\end{equation}
for all $\alpha,\,m\geq 0$, uniformly with respect to $x$.

For $\alpha = 0$, estimate (\ref{E4.7}) comes from the variation of
constant
formula and the exponential stability of the evolution operator for
(\ref{E4.6}).  For $\alpha \not= 0$, differentiate (\ref{E4.6}) with
respect to
$x\ \alpha$-times and consider the equation for $D^\alpha _xu_k$.
(\ref{E4.7})
then follows easily. We have proved the following.

\bg{thm}\label{T4.1}
Assume H1), H7), and H9) (i). Then there exists formal series
$$ u^{RSi}(x,\tau,\epsilon)=\sum_{j=0}^\infty \epsilon^j
u^{RSi}_j(x,\tau),\quad u^{RSi}_j(x,0)=\overline u_j^{Ri}(x),
$$
for $\overline\eta^{i-1}\leq x\leq \overline\eta^i,\: 1\leq i\leq r$, such
that $u^{RSi}$ formally satisfies (\ref{E4.1}). The series can be obtained
by recursively solving the system of ODEs
(\ref{E4.4})--(\ref{E4.4}${}_k$), $k\geq 1$ with the initial condition
(\ref{E4.5}). Furthermore, each $u^{RSi}_j$ is $C^m$ bounded jointly in $x$ and
$\tau$ for all $m\geq 0$.
\end{thm}

\subsection {Formal series in ${\bold (SS)^i}$ and matching of
${\bold (SS)^i}$ with ${\bold (SR)^i}$}

The position of the wave front $\eta ^i(t,\epsilon) = \displaystyle \sum
^\infty_{j=0} \epsilon ^i\eta ^i_j(t)$ depends on its initial condition
$\eta
^i(0,\epsilon) = \displaystyle \sum ^\infty_{j=0} \epsilon ^j \overline
\eta
^i_j$.  We will show in this section that $\{ \overline \eta ^i_j\}
^\infty_{j=0}$ is determined by the matching of expansions in $(SS)^i$ and
$(SR)^i$.  In the fast time variable $\tau = t/\epsilon$,
\bg{equation}\label{E4.7a}
\stackrel{*}{\eta}\!{}^i(\tau,\epsilon) = \sum^\infty_{j=0} \epsilon
^j\stackrel{*}{\eta}\!{}^i_j(\tau)\stackrel{def}{=}
\eta ^i(t,\epsilon) = \eta ^i(\epsilon \tau,\epsilon).
\end{equation}
Each $\stackrel{*}{\eta}\!{}^i_j(\tau)$ is a polynomial of degree $j$.  In
fact,
from $ \displaystyle \sum ^\infty_{j=0} \epsilon ^j
\stackrel{*}{\eta}\!{}^i_j(\tau)=
\displaystyle \sum ^\infty_{j=0} \epsilon ^j\eta ^i_j(\epsilon \tau)$,
we have
\begin{equation}\label{E4.8}
\stackrel{*}{\eta}\!{}^i_j(\tau) = \sum ^j_{\ell = 0} \eta ^{i(\ell)}_{j-\ell}
(0)
\tau ^\ell /\ell!
\end{equation}
where $(\ell)$ denotes the $\ell$-th derivative with respect to $t$.  In
particular,
\begin{equation}\label{E4.9}
\stackrel{*}{\eta}\!{}^i_j(0) = \eta ^i_j(0) = \overline \eta ^i_j.
\end{equation}

The stretched variable $\xi = (x - \displaystyle \sum ^\infty_{j=0}
\epsilon
^j\stackrel{*}{\eta}\!{}^i_j(\tau))/\epsilon$ is used to express the solution
$u^{SSi}(\xi,\tau,\epsilon)$.  When $\tau = 0$, in the new variable, the
initial data are
$$
\overline u^{Si} (\xi, \epsilon)
= \sum ^\infty_{j=0} \epsilon ^j \overline u^{Si}_j(\xi)\stackrel{def}{=}
\overline u(x,\epsilon) = \overline u(\epsilon \xi + \sum ^\infty_{j=0}
\epsilon ^j\overline \eta ^i_j, \epsilon ).
$$
Let now $\overline \eta ^i_0 = \overline \eta ^i$.  Recall the definition
of
$\overline {\overline u}^{Si}(\xi,\epsilon) = \displaystyle \sum
^\infty_{j=0}
\epsilon ^j \overline {\overline u}^{Si} (\xi)$ It is easy to see that
$$
\overline u^{Si}(\xi,\epsilon) = \overline {\overline u}^{Si} (\xi +
\sum
^\infty _{j=0} \epsilon ^j\overline \eta ^i_{j+1}, \epsilon).
$$
\begin{align}
\overline u_0(\xi) & =  \overline {\overline u}_0 (\xi + \overline \eta
^i_1),\label{E4.10}\\
\overline u_1(\xi) & =  \overline {\overline u}_1(\xi + \overline \eta
^i_1) +
\overline {\overline  u}_{0\xi} (\xi + \overline \eta ^i_1)\overline \eta
^i_2,\tag{\ref{E4.10}${}_1$}\\
&\ldots\notag\\
\overline u_k(\xi) & =  \overline {\overline u}_k(\xi + \overline u^i_1) +
\ldots + \overline {\overline u}_{0\xi} (\xi + \overline u^i_1)\overline
\eta
^i_{k+1}.
\tag{\ref{E4.10}${}_k$}
\end{align}
Observe that in (\ref{E4.10}${}_k$), the ... comprises of terms containing only
$\{\overline \eta ^i_j\}^k_{j=1}$.

Let the solution in $(SS)^i$ be
$$
u^{SSi}(\xi,\tau,\epsilon) = u^{SSi}((x -
\stackrel{*}{\eta}\!{}^i(\tau,\epsilon))/\epsilon , \tau, \epsilon) =
\sum^\infty_{j=0} \epsilon ^ju^{SSi}_j(\xi,\tau)
$$
that satisfies
\begin{equation}\label{E4.11}
u_\tau = u_{\xi\xi} + {1\over \epsilon} D_\tau
\stackrel{*}{\eta}\!{}^i(\tau,\epsilon) \cdot u_\xi + f(u, \epsilon \xi +
\stackrel{*}{\eta}\!{}^i(\tau,\epsilon),\epsilon).
\end{equation}
Observe that
\begin{equation}\label{E4.12}
{1\over \epsilon}D_\tau\stackrel{*}{\eta}\!{}^i(\tau, \epsilon) = \sum
^\infty_{j=0} \epsilon ^j \stackrel{*}{\eta}\!{}^{i(1)}_{j+1}(\tau) =
\sum ^\infty _{j=0}
\epsilon ^j ( \sum ^j_{\ell = 0} \eta ^{i(\ell +1)}_{j-\ell} (0) \tau ^\ell
/
\ell!)
\end{equation}
where (1) denotes derivatives with respect to $\tau$.  Notice that
$\stackrel{*}{\eta}\!{}^{i(1)}_{j+1}(\tau)$ is a polynomial of degree $j$.
Expanding (\ref{E4.11}) and using (\ref{E4.8}) and (\ref{E4.12}), we have a
system for $\sum ^\infty_{j=0} \epsilon ^ju^{SSi}_j$.  (Recall $\eta
^{i(1)}_0(0) = V^i(\overline \eta ^i)$ from (\ref{E3.11}), and $V^i(\overline
\eta ^i) = 0$ for $i=0,r$.)
\begin{equation}\label{E4.13}
u_{0\tau}  =  u_{0\xi\xi} + V^i(\overline \eta ^i)u_{0\xi} +
f_0(u_0,\overline \eta ^i),
\end{equation}
\bg{equation}\tag{\ref{E4.13}${}_1$}
\bg{aligned}
u_{1\tau} & =  u_{1\xi\xi} + V^i(\overline \eta ^i)u_{1\xi} +
f_{0u}(u_0,\overline \eta ^i)u_1\\
& +  \stackrel{*}{\eta}\!{}^{i(1)}_2(\tau) u_{0\xi} + f_{0x} (u_0,\overline
\eta ^i)
(\xi + \stackrel{*}{\eta}\!{}^i_1(\tau))+f_1(u_0,\overline \eta ^i),
\end{aligned}
\end{equation}
$$ \ldots $$
\bg{equation}\tag{\ref{E4.13}${}_k$}
\bg{aligned}
u_{k\tau} & =  u_{k\xi\xi} + V^i(\overline \eta ^i)u_{k\xi} +
f_{0u}(u_0,\overline \eta ^i)u_k\\
& + \sum ^k_{j=1} \stackrel{*}{\eta}\!{}^{i(1)}_{j+1} (\tau) u_{k-j,\xi}+\sum
C_{\alpha \beta \gamma \delta}D^{|\alpha|}_uD^{|\beta| + \gamma}_x f_\delta
(u_0 ,
\overline \eta ^i) u^\alpha  \eta ^\beta \xi^ \gamma .
\end{aligned}
\end{equation}
Here $\alpha = (\alpha _1, \ldots, \alpha_{k-1}),\, u^\alpha = u^{\alpha
_1}_1
\ldots u^{\alpha _{k-1}}_{k-1},\, \beta = (\beta _1, \ldots , \beta _k),\,
\eta ^\beta = \stackrel{*}{\eta}\!{}^{\beta}_1\ldots
\stackrel{*}{\eta}\!{}^{\beta_k}_k,\,
\gamma, \delta \geq 0$ are integers, $\delta + \gamma + \displaystyle \sum
^{k-1}_{j=1} (\alpha _j + \beta _j) \cdot j + k\beta _k = k$, $C_{\alpha
\beta \gamma \delta}$ is a constant.  The initial conditions for $u_0, u_1,
\ldots,
u_k$ are given in (\ref{E4.10})--(\ref{E4.10}${}_k$), $k\geq 1$.

The existence of local solutions for (\ref{E4.13})--(\ref{E4.13}${}_k$),
$k\geq 1$, follows from the theory of
abstract parabolic equations and analytic semigroup.
Let $Au=u_{\xi\xi}+V^i(\overline\eta^i) u_\xi$.
Let $1 \leq i \leq r - 1$ first. Consider (\ref{E4.13}) in
$X = B_{\Bbb R}(w)$. Then $A$ is sectorial with $D_A = B^2_{\Bbb R}(w)$.
Since $u_0(\xi,0) = \overline
u^i_0(\xi) \in D_A$, (\ref{E4.13}) admits a unique solution
\begin{equation}\label{E4.14a}
u \in C^1 ([0,t_0]: X ) \cap C([0,t_0]: D_A)
\end{equation}
for some $t_0 > 0$.  If $i = 0$ or $r$, the same conclusion hold but
$X =B_{{\Bbb R}^\pm} (w)$ and $D_A= B_{{\Bbb R}^\pm} (w) \cap (BC)$.
However, due to H9), the solution
$u^{SSi}_0(\xi,\tau)$ exists for any $t_0 > 0$ and approaches $q^i(\xi + c,
\overline \eta ^i)$ as $\tau \to \infty$, for some constant $c \in {\Bbb
R}$.
When $1 \leq i \leq r - 1$, (\ref{E4.13}${}_k$), $k\geq 1$, is considered in
$X = E_{\Bbb R}(w)$
and $A$ is sectorial with $D_A
= E^2_{\Bbb R}(w)$.  When $i = 0,r,\, X = E_{{\Bbb R}^\pm} (w)$ and $D_A =
E^2_{{\Bbb R}^\pm} \cap (BC)$.  Since $u_k(\xi,0) = \overline u^i_k(\xi) \in
D_A$, and it can be seen by induction that
the right hand side of (\ref{E4.13}${}_k$) is in
$D_A(\frac{1}{2})$, (\ref{E4.13}${}_k$) admits a solution $u$ that
satisfies (\ref{E4.14a}).
Since the evolution operator of (\ref{E4.13}${}_k$)
is not asymptotically stable, and the nonhomogeneous terms are of
$O((1 + \tau )^k)$ in $D_A(\frac{1}{2})$, in general,
we can show $|u_k|_{E^2(1 + |\xi|^k)} = O((1 + \tau )^{k+1})$.  However,
better result can be obtained by considering the matching of $(SS)^i$ and
$(SR)^i$.  In our case, $u_k(\xi,\tau)$ exists for all $\tau \in [0,\infty)$
and
satisfies $\vert D^\alpha_\tau u_k(\xi,\tau) \vert+ \vert D^\beta_\xi
u_k(\xi,\tau) \vert \leq C(1 + |\xi|^k + \tau ^k)$ for $\alpha\leq 1,\:
\beta\leq 2$.

For the purpose of matching we expand $u^{SRi}(\xi,t,\epsilon) =
\displaystyle
\sum ^\infty_{j=0} \epsilon ^ju^{SRi}_j(\xi,t)$ in the fast time $\tau =
t/\epsilon$.
$$
u^{SRi}(\xi,\epsilon\tau,\epsilon) =
\stackrel{*}{u}\!{}^{SRi}(\xi,\tau,\epsilon)
= \sum ^\infty_{j=0} \epsilon ^j \stackrel{*}{u}\!{}^{SRi}_j (\xi,\tau).
$$
Since $u(x,t,\epsilon) = \stackrel{*}{u}\!{}^{SRi} ((x -
\stackrel{*}{\eta}(\tau,\epsilon))/\epsilon , \tau, \epsilon),\
\stackrel{*}{u}\!{}^{SRi} (\xi,\tau,\epsilon)$ formally satisfies the same
equations as (\ref{E4.11}),  therefore $\displaystyle \sum ^\infty_{j=0}
\epsilon ^j\stackrel{*}{u}\!{}^{SRi}_j(\xi,\tau)$ formally satisfy system
(\ref{E4.13}), (\ref{E4.13}${}_1$), ..., (\ref{E4.13}${}_k$), ... .
Also from
$$
\sum ^\infty_{j=0} \epsilon ^ju^{SRi}_j (\xi, \epsilon \tau) = \sum
^\infty_{j=0} \epsilon ^j\stackrel{*}{u}\!{} ^{SRi}_j(\xi,\tau),
$$
and $D^\ell_tu^{SRi}_j(\xi,t) \in E^m(1 + |\xi|^j)$ for all $\ell,m \geq
0,\
\stackrel{*}{u}\!{}^{SRi}_j$ is a polynomial in $\tau$, in the form
\begin{equation}\label{E4.14}
\stackrel{*}{u}\!{}^{SRi}_j = \sum ^j_{\ell=0} \stackrel{*}{u}_{j\ell}, \mbox {
with }
| \stackrel{*}{u}_{j\ell}| _{E^m(1 + |\xi|^\ell)} \leq C\tau ^{j-\ell},
\mbox {
for all } m \geq 0.
\end{equation}
In particular, $\stackrel{*}{u}\!{}^{SRi}_j \in O ( 1 + |\xi|^j + \tau ^j)$.

Observe here that $\{u^{SRi}_j\}_o^\infty$ has not be determined since
$\{\overline\eta^i_j\}_1^\infty$ is still unknown. But knowing
$\{\overline\eta^i_j\}_0^k$ suffices to compute $\{u^{SRi}_j\}_0^k$ and
$\{\eta^i_j(t)\}_0^k$.

We now prove by induction that by successively choosing $\{\overline \eta
^i_j\}^\infty_{j=1}$, ($\overline \eta ^i_{j+1}$ affects initial condition
$\overline u_j(\xi)$, cf. (\ref{E4.10}${}_j$)),
system (\ref{E4.13})--(\ref{E4.13}${}_j$) has  a unique
solution $\{u^{SSi}_j\}^\infty_{j=0}$ such that
\begin{equation}\label{E4.15}
u^{SSi}_j = \sum ^k_{j=0} u_{j\ell} \mbox { with }
\|u_{j\ell}\|_{E^2(1+|\xi|^\ell)} \leq C(1 + \tau ^{j-\ell}),
\end{equation}
and
\begin{equation}\label{E4.16}
|u_{j\ell} - \stackrel{*}{u}_{j\ell}|_{E^2(1 + |\xi|^\ell)} \leq C(1 + \tau
^{j-\ell}) e^{-\gamma \tau}.
\end{equation}

First, if $1\leq i\leq r-1$, (\ref{E4.13}) has a stable stationary
solution $q^i(\xi,\overline \eta
^i)$ that attracts $\overline {\overline u}_0$ modulo spatial shifts.  cf
H9).  Thus, there exists a unique $\overline \eta ^i_1$ such that with
$u_0(\xi,0) = \overline u_0(\xi) = \overline {\overline u}(\xi + \overline
\eta ^i_1)$,
\begin{equation}\label{E4.17}
\| u^{SSi}_0 ( \cdot , \tau) - q^i(\cdot , \overline \eta^i)\|_{B^2_{\Bbb
R}(w)} \leq Ce^{-\gamma \tau}, \ \tau \geq 0,
\end{equation}
where $w(\xi) = e^{-\gamma|\xi|}$ and $\gamma > 0$. If $i=0,r$,
$q^i(\xi,\overline \eta
^i)$ is stable. We choose $\overline\eta ^i_1 =0$, (\ref{E4.17}) is still
valid. Now (\ref{E4.17}) is even stronger than
(\ref{E4.16}), $j=\ell=0$.  The extra control of the rate of approaching
$u_0(\pm \infty,\tau)$ as $\xi \to \pm \infty$ will be used in \S 6.
At this point, $\{u^{SRi}_j\}_{j=0}^1$ and $\{\eta^i_j\}_{j=0}^1$ have
also been determined.

Assume that $\{u^{SSi}_j\}^{k-1}_{j=0}$ and $\{\overline \eta
^i_j\}^k_{j=0}$
have been determined and (\ref{E4.15}), (\ref{E4.16}),\, $0 \leq j \leq k-1$
are
satisfied.  Therefore, $\eta ^i_j(t),\, t \geq 0,\, 0 \leq j \leq k$ and all
their derivatives at $t = 0$ are determined as well as
$\{u^{SRi}_j\}_0^k$ and $\{\stackrel{*}{u}\!{}^{SRi}_j \}_0^k$.
{}From (\ref{E4.8}) and
(\ref{E4.12}),\,
$\stackrel{*}{\eta}\!{}^i_j(\tau)$ and $\stackrel{*}{\eta}\!{}^{i(1)}_{j+1}
 (\tau),\, 0 \leq j \leq k$ are determined,
which will be used in (\ref{E4.13}${}_k$).  We now
rewrite (\ref{E4.13}${}_k$), which is satisfied by $u^{SSi}_k$ and
$\stackrel{*}{u}\!{}^{SRi}_k$ as the followings:
$$
\begin{array}{l}
u_{k\tau} - u_{k\xi\xi} - V^i(\overline \eta ^i) u_{k\xi} -
f_{ou}(u_0,\overline \eta ^i)u_k  = g_k(u_0, \ldots , u_{k-1}),\\
\stackrel{*}{u}_{kt} - \stackrel{*}{u}_{k\xi\xi} - V^i(\overline \eta
^i)\stackrel{*}{u}_{k\xi} - f_{ou}(\stackrel{*}{u}_0,\overline \eta
^i)\stackrel{*}{u}_k  = g_k(\stackrel{*}{u}_0, \ldots ,
\stackrel{*}{u}_{k-1}).
\end{array}
$$
Since $g_k(u_0,\ldots,u_{k-1})$ is a polynomial in $(u_1, \ldots , u_{k-1})$
and $(u_{1\xi},\ldots,u_{k-1,\xi})$, using
(\ref{E4.14})--(\ref{E4.16}) we can verify that
$$
g_k(u_0, \ldots , u_{k-1}) - g_k (\stackrel{*}{u}_0, \ldots ,
\stackrel{*}{u}_{k-1}) = \sum ^k_{j=0} g_{kj}
$$
with $\|g_{kj}\|_{E^1(1 + \|\xi\|^j)} \leq C( 1 + \tau ^{k-j}) e^{-\gamma
\tau}$.  Let $\Delta u = u_k - \stackrel{*}{u}_k$.

Denote
$$
L^i u = u_{\xi\xi} + V^i(\eta ^i) u + f_{ou}(u^i_0,\overline \eta ^i)u
$$
\begin{equation}\label{E4.18}
\begin{aligned}
 \Delta u_\tau - L^i \Delta u & =  \sum ^k_{j=0}
\left [ f_{0u}(u_0,\overline \eta ^i) -
f_{0u}(\stackrel{*}{u}_0,\overline \eta ^i)\right ] \stackrel {*}{u}_{kj} +
\sum ^k_{j=0} g_{kj} \\
& =  \sum ^k_{j=0} G_{kj}.
\end{aligned}
\end{equation}
At this point all the terms in the right hand side of (\ref{E4.18}) are
known. Observe that the initial data for $\Delta u$ has the form
 $u_k(\xi,0) - \stackrel{*}{u}_k(\xi,0) = \varphi (\xi) + \overline
{\overline u}_{0\xi} (\xi + \overline \eta ^i_1)\overline \eta ^i_{k+1}$,
where
$\varphi \in E^2(1 + |\xi|^k)$.  Recall that  $\overline\eta^0_0=a,\:
\overline\eta_o^r=b,\:\overline \eta ^i_j = 0$ for all $j
\geq 1$ if $i = 0,r$.

\begin{lem}\label{L4.3} (i) Let $1 \leq i \leq r - 1$.  Consider
\begin{equation}\label{E4.19}
\begin{aligned}
u_\tau & =  L^i u + h,\\
u(\xi,0) & =  \varphi (\xi) + \overline {\overline u}_{0\xi} (\xi + \overline
\eta ^i_1)\eta,
\end{aligned}
\end{equation}
where $h : {\Bbb R}^+ \to E^1_{\Bbb R}(1 + |\xi|^j)$ is continuous with
$|h(\tau)|_{E^1(1+|\xi|^j)} \leq C(1 + \tau ^\ell) e^{-\gamma \tau}$ and
$\varphi \in E^2_{\Bbb R}(1 + |\xi|^j)$.  Then there exists a unique $\eta
\in
{\Bbb R}$ such that there exists a unique solution $u$ to (\ref{E4.19})
with $|u(\tau)|_{E_{\Bbb R}^2(1 + |\xi|^j)} \leq C(1 + \tau
^\ell) e^{-\gamma\tau}$.

(ii)  Let $i = 0$ or $r$.  Consider (\ref{E4.19}) with $\overline \eta ^i_1
= \eta = 0$, where $h:{\Bbb R}^+ \to E^1_{{\Bbb R}^\pm}(1 + |\xi|^j)$ is
continuous with $|h(\tau)|_{E^1_{{\Bbb R}^\pm}(1+|\xi|^j)} \leq C(1 + \tau
^\ell) e^{-\gamma \tau}$ and $\varphi \in E^2_{{\Bbb R}^\pm} (1 + |\xi|^j)
\cap (BC)$.  Then there exists a unique solution $u$ to (\ref{E4.19}) such
that
$u:{\Bbb R}^+ \to E^2_{{\Bbb R}^\pm} (1 + |\xi|^j)\cap (BC)$ is continuous and
$|u(\tau)|_{E^2_{{\Bbb R}^\pm}(1 + |\xi|^j)} \leq C(1 +
\tau^\ell)e^{-\gamma
\tau}$.
\end{lem}

We now write $\Delta u_k = \displaystyle \sum ^k_{j=0} \Delta u_{kj}$ where
$\Delta u_{kj}$ satisfies
\begin{equation}\label{E4.20}
u_\tau - L^iu   =  [f_{ou}(u^i_0,\overline \eta ^i) -
f_{0u}(\stackrel{*}{u}\!{}^i_0,\overline \eta ^i)]\stackrel{*}{u}_{kj} + g_{kj}
=
G_{kj}.
\end{equation}
\begin{equation}\label{E4.21}
u(\xi,0)  =  \left \{ \begin{array}{rl}
\overline {\overline u}_{0\xi} (\xi + \overline \eta^i_1)C_j, & \mbox { if
} 0
\leq j \leq k - 1,\\
\varphi (\xi) + \overline {\overline u}_{0\xi}(\xi + \overline \eta ^i_1)
C_k,
& \mbox { if } j = k.\end{array}
\right .
\end{equation}

Let $1 \leq i \leq r -1$ first.  Since $\varphi \in E^2(1 + |\xi|^k)$ and
$|G_{kj}|_{E^1_{\Bbb R}(1+|\xi|^j)} \leq C(1 + \tau
^{k-j})e^{-\gamma\tau}$,
from \lemref{L4.3}, (i), there exists a unique $C_j, 0 \leq j \leq k$ such
that the unique solution $\Delta u_{kj}$ of (\ref{E4.20}), (\ref{E4.21})
satisfies
\begin{equation}\label{E4.22}
|\Delta u_{kj} |_{E^2_{\Bbb R}(1 + |\xi|^j)} \leq C(1 + \tau
^{k-j})e^{-\gamma \tau}.
\end{equation}

Let $i = 0$ or $r$ next.  Then $\varphi (\xi) = \overline {\overline
u}_k(\xi) =
\overline u_k(\xi) \in E^2_{{\Bbb R}^\pm} (1 + |\xi|)^k) \cap (BC)$ and
$|G_{kj}|_{E^1_{{\Bbb R}^\pm}(1 + |\xi|^j)}  \leq C(1 + \tau
^{k-j})e^{-\gamma
\tau}$ and $C_j = 0,\: 0 \leq j \leq k$.  From \lemref{L4.3}, (ii), the unique
solution $\Delta u_{kj}$ of (\ref{E4.20}), (\ref{E4.21}) satisfies the
Neumann boundary condition at $\xi=0$ and
\begin{equation}
\label{E4.23}
|\Delta u_{kj}|_{E^2_{{\Bbb R}^\pm}(1 + |\xi|^j)} \leq C( 1 + \tau
^{k-j})e^{-\gamma \tau}
\end{equation}
In all the cases let $\overline\eta^i_{j+1}=\sum_0^k c_j$ and
$u^i_{kj} = \Delta u_{kj} + \stackrel{*}{u}_{kj}$.  From (\ref{E4.14}),
we
have (\ref{E4.15}) for $j = k$.  Thus, (\ref{E4.15}) and (\ref{E4.16}) have
been proved by the induction.

We summarize the results in
\bg{thm}\label{T4.2}
Assume H1)--H5). For each $0\leq i\leq r$ assume the initial data
$\overline {\overline u}^{Si}(\xi,\epsilon)=\sum\epsilon^j \overline
{\overline u}^{Si}_j(\xi)$ satisfies H7) and H9) (ii). Then there exist
formal series
\bg{equation}\label{E4.24}
u^{SSi}(\xi,\tau,\epsilon)=\sum\epsilon^j u^{SSi}_j(\xi,\tau),\quad
u^{SSi}_j(\xi,0)=\overline u^{Si}_j(\xi),
\end{equation}
\bg{equation}\label{E4.25}
\sum\epsilon^j \overline\eta^i_j, \quad \overline\eta^i_0 =\overline \eta^i.
\end{equation}
Here $\overline\eta^i$ is determine by the layer position of the initial
data. (\ref{E4.24}) formally satisfies (\ref{E4.11}) where
$\stackrel{*}{\eta}\!{}^i(\tau,\epsilon)=\eta^i(\epsilon\tau,\epsilon)$
is the expansion of the layer position in the variable $\tau=t/\epsilon$,
see (\ref{E4.7a}). Furthermore (\ref{E4.24}) is recursively determined
by (\ref{E4.13})--(\ref{E4.13}${}_k$), $k\geq 1$, with the initial conditions
(\ref{E4.10})--(\ref{E4.10}${}_k$), $k\geq 1$, in the space $u^{SSi}_j\in
E^2(1+|\xi|^j)$, where $\overline\eta^i_j,\:j\geq 1$ is chosen such that
for any integers $\alpha\leq 2,\:\beta\leq 1$,
\bg{align*}
\vert \partial^\alpha_\xi u^{SSi}_j\vert + \vert \partial^\beta_\tau u^{SSi}_j
\vert
&\leq C (1+|\xi|^j+\tau^j),\\
\vert u^{SSi}_j - \stackrel{*}{u}\!{}^{SRi}_j \vert
&\leq C ((1+|\xi|^j+\tau^j)e^{-\gamma\tau},
\end{align*}
for some $\gamma > 0$.
\end{thm}

\endinput


\section{Matching of layer solutions, construction of pseudo solutions}
\subsection{The matching of $(RS)^i$ and $(RR)^i$}
\bg{thm}\label{T5.1}
There exists $\gamma > 0$ such that for any integers
$\alpha,\,\beta,\,j\geq 0$,
$$
 \sup_{\tau\geq 0}\left\{\vert \partial_\tau^\alpha \partial_x^\beta
 (u_j^{RSi}(x,\tau) - u_j^{RRi}(x)) \vert\right \}\leq C_{\alpha\beta
j}e^{-\gamma\tau},
$$
uniformly for all $x\in[\overline\eta^{i-1},\,\overline\eta^i], \:
1\leq i\leq r$.
\end{thm}

When $j=0$, \thmref{T5.1} is a consequence of the exponential stability of
$p^i(x), \: 1\leq i\leq r$, as a stationary solution to (\ref{E4.4}).
By induction, we can prove that the nonhomogeneous term and the
coefficient $f_{0u}(u_0,x)$ in (\ref{E4.4}${}_k$) approach the
corresponding terms in (\ref{E3.8}${}_k$) exponentially as $\tau\to\infty$.
The desired result than follows easily. Details are omitted.

\subsection{The Matching of $(SS)^i$ and $(SR)^i$}
The matching of expansions in $(SS)^i$ and $(SR)^i$ has been obtained in
\S 5.3. For convenience, we state the result in the
following
\bg{thm}\label{T5.2}
Let $\Delta u^i_j = u_j^{SSi} - \stackrel{*}{u}\!{}^{SRi}_j$, where
$\sum_0^\infty\epsilon^j \stackrel{*}{u}\!{}^{SRi}_j$ is the expansion of
$\sum_0^\infty\epsilon^j u_j^{SRi}(\xi,t)$ in the variable $\tau=t/\epsilon$.
Then there exists $\gamma>0$ such that for any nonnegative
integers $\alpha\leq 2,\: \beta\leq 1$,
$$
 \left\{\vert \partial^\alpha_\xi \Delta u^i_j \vert +\vert
 \partial^\beta_\tau \Delta u^i_j \vert\right\} \leq C_j(1+|\xi|^j+\tau^j)
 e^{-\gamma \tau}.
$$
\end{thm}

\subsection{The matching of $(SR)^i$ with $(RR)^i$ and $(RR)^{i+1}$.}

Let the inner expansion of the outer formal solutions in $(RR)^i$ be
\begin{equation}\label{E5.1}
\bg{aligned}
\sum ^\infty_{j=0} \epsilon ^j \tilde u^{RRi}_{j,1} (\xi, t) & =   \sum
^\infty_{j=0} \epsilon ^ju^{RRi}_j (\epsilon \xi + \sum ^\infty _{\ell = 0}
\epsilon ^\ell \eta ^i_\ell (t)),\quad  1 \leq i \leq r,\\
\sum ^\infty_{j=0} \epsilon ^i\tilde u^{RRi}_{j,2} (\xi, t) & =  \sum
^\infty_{j=0} \epsilon ^ju^{RR(i+1)}_j(\epsilon \xi + \sum ^\infty_{\ell =
0} \epsilon ^\ell \eta ^i_\ell (t)),\quad  0 \leq i \leq r - 1.
\end{aligned}
\end{equation}
Recall that $\eta ^0_0(t) = a,\, \eta ^r_0(t) = b$ and $\eta ^0_\ell (t) =
\eta
^r_\ell(t) = 0$ for all $\ell \geq 1$.  From (\ref{E5.1}), we find that
$\tilde u^{RRi}_{j,\nu}, \nu = 1,2$ is a polynomial in $\xi$  of degree  $j$,
with its
coefficients depending smoothly on $\eta ^i_\ell(t),\, \ell \leq j$.  Thus,
$$
\tilde u^{RRi}_{k,\nu} \in E^m_{\Bbb R}(1 + |\xi|^k),\ \nu = 1,2
$$
with the norm bounded uniformly with respect to $t \geq 0$.

Concerning the matching of $(SR)^i$ and $(RR)^i$, we want to show that
there exists $\gamma>0$ such that the following estimates hold.
\begin{equation}\label{E5.2}
\bg{aligned}
\vert u^{SRi}_j(\xi,t)  -  \tilde u^{RRi}_{j,1} (\xi,t)\vert_{E^m_{{\Bbb
R}^-}((1 _+ |\xi|^j)e^{-\gamma|\xi|})} &\leq C_{mj},\quad 1\leq i\leq r,\\
\vert u^{SRi}_j(\xi,t)  -  \tilde u^{RRi}_{j,2}(\xi,t)\vert_{E^m_{{\Bbb R}^+}
((1 + \xi^j)e^{-\gamma \xi})} &\leq C_{mj},\quad 0\leq i\leq r-1,
\end{aligned}
\end{equation}
for all $m,\,j \geq 0$ uniformly uniformly with
respect to $t \geq 0$ in the weighted norms.  Only the proof of the second
estimate, $0\leq i\leq r-1$, will be presented since the proof of the first
is similar.

For $j = 0$, it is clear that
$u^{SRi}_0(\xi,t) = q^i(\xi,\eta ^i_0(t))$ approaches $p^{i+1} (\eta
^i_0(t))
= \tilde u^{RRi}_{0,2}(t)$ as a $\xi \to + \infty$.  The rate  is bounded
by
$Ce^{-\gamma \xi}$.  Since $x = \eta ^i_0(t)$ is in a compact subinterval of
$O^i$, and the solution depends continuously with respect to $x$, therefore,
$C$ does not depend on $t \geq 0$.

Assume (\ref{E5.2}) is valid for all $j \leq k -1$.  Notice that
$\displaystyle
\sum ^\infty_{j=0} \epsilon ^j\tilde u^{RRi}_{j,2} (\xi,t)$ formally
satisfies
$$
\epsilon \tilde u_t = \tilde u_{\xi\xi} + D_t \eta (t,\epsilon) \tilde
u_{\xi}
+ f(\tilde u, \epsilon \xi + \eta (t,\epsilon),\epsilon)
$$
This has the same form as (\ref{E3.9a}).  Expanding in power series of
$\epsilon$, the equations for $\{\tilde u^{RRi}_{j,2}\}^\infty_{j=0}$ are
precisely (\ref{E3.10}), (\ref{E3.10}${}_1$), ..., (\ref{E3.10}${}_k$),
..., and can be rewritten in a form as (\ref{E3.12}) if $j\geq 1$,
$$
\tilde u_{k\xi\xi} + V^i(x) \tilde u_{k\xi} + f_{0u}(\tilde u_0,x)\tilde
u_{k} +
V_k\tilde u_{0\xi} = h_k(\xi,\tilde u_0,...\tilde u_{k-1},x_1,...,x_k),
$$
where $x = \eta ^i_0(t), ..., x_j = \eta ^i_j(t)$.

For $1\leq i \leq r-1$, $\Delta u_k\stackrel{def}{=}u_k - \tilde u_k$ satisfies
\bg{equation}\label{E5.3}
\begin{split}
&\Delta u_{k\xi\xi} + V^i(x) \Delta u_{k\xi} + f_{0u} (\tilde
u_0,x)\Delta u_k \\
=& h_k(\xi,u_0, ... u_{k-1}, x, ..., x_k) - h_k(\xi, \tilde u_0, ... ,
\tilde
u_{k-1}, x, ... , x_k)\\
+& [f_{0u}(\tilde u_0,x) - f_{0u}(q^i(\xi,x),x)]u_k + V_k[\tilde u_{0\xi} -
q^i_\xi(\xi,x)].
\end{split}
\end{equation}
Eq. (\ref{E5.3}) is considered for $\xi \in {\Bbb R}$.
The right hand side is in $E^m(1 + |\xi|^k)$ and $E^m((1 +
|\xi|^k)e^{-\gamma\xi})$, with the norms bounded uniformly with respect to
$t\geq 0$ due to the induction assumption.  Recall that
$\tilde
u_0 = p^{i+1}$ and $L_{p^{i+1}}$ has an exponential dichotomy on ${\Bbb R}$.
{}From
\lemref{LB.2} (i), (\ref{E5.3}) has a unique solution $U_k$ that is in both
$E^{m+2}(1
+|\xi|^k)$ and $E^{m+2}((1 + |\xi|^k)e^{-\gamma\xi})$.  However, it is known
that
$u_k, \tilde u_k$, therefore $\Delta u_k \in E^{m+2}(1 + |\xi|^k)$.  We then
have
$\Delta u_k = U_k \in E^{m+2}((1 + |\xi|^k)e^{-\gamma \xi})$.

For $i = 0$, $\Delta u_k\stackrel{def}{=}u_k - \tilde u_k$ satisfies
\bg{equation}\label{E5.3a}
\begin{split}
&\Delta u_{k\xi\xi} + V^i(x^i) \Delta u_{k\xi} + f_{0u} (q^i(\xi,x^i),x^i)
\Delta u_k \\
=& h_k(\xi,u_0, ... u_{k-1}, x, ..., x_k) - h_k(\xi, \tilde u_0,
... ,
\tilde u_{k-1}, x, ... , x_k)\\
+& [f_{0u}(\tilde u_0,x) - f_{0u}(q^i(\xi,x^i),x^i)]\tilde u_k + V_k[\tilde
u_{0\xi} - q^i_\xi(\xi,x^i)],
\end{split}
\end{equation}
\bg{equation*}\tag{\ref{E5.3a}${}_1$}
\Delta u_{k\xi}(0,t)= - \tilde u_{k\xi}(0,t),
\end{equation*}
where $x=x^i=a$, $x_1=\ldots=x_{k-1}=0$.
Eq. (\ref{E5.3a}) is considered in ${\Bbb R}^+$. Since
$\dis \sup_{t\geq 0}|-\tilde u_{k\xi}(0,t)|<C$,  from
\lemref{LB.2} (ii), (\ref{E5.3a}) and (\ref{E5.3a}${}_1$) have a unique
solution $U_k$ that is in
$E^{m+2}_{{\Bbb R}^+} (1 + |\xi|^k)$ and $E^{m+2}_{{\Bbb R}^+}((1 +
|\xi|^k)e^{-\gamma \xi})$.  However, we know that $\Delta u_k$ is a
solution of (\ref{E5.3})and (\ref{E5.3a}${}_1$) and is in $E^{m+2}
(1 + |\xi|^k)$. Thus
$$
 \Delta u_k=U_k\in E^{m+2}_{{\Bbb R}^+}((1 + |\xi|^k)e^{-\gamma \xi}).
$$

Since equation (\ref{E5.3}) depends continuously on $x, \ldots x_k$ that
is in a compact subset of ${\Bbb R}^{k+1}$, the norm of $\Delta u$ in such
space is uniformly bounded with respect to $t\geq 0$.
The second estimate of (\ref{E5.2}) has been  been proved by induction.

\bg{thm}\label{T5.3}
Let $\Delta u_j^i= u_j^{SRi}-\tilde u_{j,1}^{RRi},\:1\leq i\leq r$ or
$u_j^{SRi}-\tilde u_{j,2}^{RRi},\:0\leq i\leq r-1$, where
$\tilde u_{j,1}^{RRi}$ and
$\tilde u_{j,2}^{RRi}$ are defined in (\ref{E5.1}). Then there exists
$\gamma>0$ such that for all integers $\alpha,\,\beta,j \geq 0$,
$$
 \vert\partial^\alpha_\xi\Delta u^i_j\vert + \vert\partial^\beta_t \Delta
 u^i_j\vert \leq C_{\alpha\beta j} (1+|xi|^j)e^{-\gamma |\xi|}
$$
uniformly with respect to $t\geq 0$.
\end{thm}

\subsection{Matching of ${(SS)^i}$ with ${(RS)^i}$ and ${(RS)^{i+1}}$.}

We use the inner variable $\xi = (x -
\stackrel{*}{\eta}(\tau,\epsilon))/\epsilon$ to expand the outer solutions
\bg{equation}\label{E5.4a}
\begin{aligned}
\sum ^\infty_{j=0} \epsilon ^ju^{RSi}_j(\epsilon \xi +
\stackrel{*}{\eta}\!{}^i(\tau,\epsilon),\tau) &= \sum ^\infty_{j=0}\epsilon
^j\tilde u^{RSi}_{j,1}(\xi,\tau),\ 1 \leq i \leq r, \\
\sum ^\infty_{j=0} \epsilon ^ju^{RS,i+1}_j(\epsilon \xi +
\stackrel{*}{\eta}\!{}^i(\tau,\epsilon),\tau) &= \sum ^\infty_{j=0} \epsilon
^j\tilde u^{RSi}_{j,2}(\xi,\tau),\ 0 \leq i \leq r-1.
\end{aligned}
\end{equation}
Both expansions formally satisfy (\ref{E4.11}) as does
$u^{SSi}(\xi,\tau,\epsilon)$.
Therefore, both satisfy (\ref{E4.13})-- (\ref{E4.13}${}_k$), $k\geq 1$,
just as
$\dis \sum ^\infty_{j=0}\epsilon ^ju^{SSi}_j$.  It suffices to show the
matching of expansions in $(SS)^i$ and $(RS)^{i+1},\:0\leq i\leq r-1$.
Let the indices be dropped
so that $\tilde u_j$ denotes $\tilde u^{RSi}_{j,2}$.  Then
\begin{align}
\tilde u_{0\tau} & =  f_0(\tilde u_0,\overline \eta ^i), \quad \tilde u(0)
= \overline u^{R,i+1}_0(\overline \eta ^i)\label{E5.4}\\
u^{SSi}_{0\tau} & =  u^{SSi}_{0\xi\xi} + V^i(\overline \eta ^i)
u^{SSi}_{0\xi} + f_0 (u^{SSi}_0, \overline \eta ^i)\label{E5.5}
\end{align}
Let $\xi \to + \infty$ in (\ref{E5.5}).  Since $u_0 \in B^2_{R^+} (e^{-\gamma
\xi})$,\, $u_{0\xi}$ and $u_{0\xi\xi} \to 0$ as  $\xi \to \infty$.  We  have
$u_{0\tau}(+ \infty,\tau) = f_0(u_0(+ \infty,\tau), \overline \eta ^i)$
with
$u_0(\infty,0)  = \overline {\overline u}_0(\infty) = \overline
u^{R,i+1}_0(\overline \eta ^i)$.  This is the same as (\ref{E5.4}).
Therefore
$$
u^{SSi}_0(\infty,\tau) = \tilde u_0(\overline \eta ^i,\tau),\quad \tau \geq 0.
$$
Since $u^{SSi}_0: {\Bbb R}^+ \to B^2_{\Bbb R}(e^{-\gamma \xi})$ is bounded,
we
have
\begin{equation}\label{E5.6}
\vert u^{SSi}_0(\cdot ,\tau) - \tilde u_0(\cdot , \tau) \vert _{E^2_{\Bbb
R^+}(e^{-\gamma\xi})} \leq C
\end{equation}
uniformly with respect to $\tau \geq 0$.

Let $\Delta u_j=u_j^{SSi} - \tilde u_{j,2}^{RSi}$.
We now show by induction that
\bg{equation}\label{E5.7}
\Delta u_j = \sum ^j_{i=0} \Delta u_{j\ell}, \mbox { with } |\Delta
u_{j\ell}
|_{E^2_{\Bbb R^+}((1 + \xi^\ell)e^{-\gamma \xi})} \leq C(1 + \tau
^{j-\ell}).
\end{equation}
Suppose (\ref{E5.7}) has been proved for $0 \leq j \leq k - 1$.  For
$1\leq i\leq r-1$, $\Delta u_k$ satisfies
\begin{align}
\Delta u_{k\tau} & = \Delta u_{k\xi\xi} + V^i(\overline \eta ^i) \Delta
u_{k\xi} + f_{0u}(\tilde u_0,\overline \eta ^i)\Delta u_k\notag\\
& +  [f_{0u}(u^{SSi}_0,\overline \eta ^i) - f_{0u}(\tilde u_0,\overline
\eta ^i)] u^{SSi}_k\label{E5.8}\\
& + g_k(u_0, \ldots , u_{k-1}) - g_k(\tilde u_0, \ldots , \tilde u_{k-1}).
\notag
\end{align}
For $i=0$, $\Delta u_k$ satisfies the equations
\begin{align}
\Delta u_{k\tau} & = \Delta u_{k\xi\xi} + V^i(\overline \eta ^i)
\Delta
u_{k\xi} + f_{0u}(u_0^{SSi},\overline \eta ^i)\Delta u_k\notag\\
& +  [f_{0u}(u^{SSi}_0,\overline \eta ^i) - f_{0u}(\tilde
u_0,\overline \eta ^i)]\tilde u_k\label{E5.8a}\\
& + g_k(u_0, \ldots , u_{k-1}) - g_k(\tilde u_0, \ldots , \tilde
u_{k-1}),
\notag
\end{align}
\bg{equation*}\tag{\ref{E5.8a}${}_1$}
\Delta u_{k\xi}(0,\tau)= -\tilde u^{RSi}_{j,2}(0,\tau),\quad \text{at
}\xi=0,
\end{equation*}
where $V^i=0$, $\overline\eta^i=a$.
Observe that $g_k$ is a polynomial on $u_1,  \ldots , u_{k-1}, u_{1\xi} ,
\ldots$, $u_{k-1,\xi}$ and $\stackrel{*}{\eta}\!{}^i_1, \ldots ,
\stackrel{*}{\eta}\!{}^i_k$.  By the induction assumption, the
nonhomogeneous term of (\ref{E5.8}) has the form
\bg{equation}\label{E5.9}
G_k = \sum ^k_{j=0} G_{kj},\; \mbox { with } |G_{kj}|_{E^1_{\Bbb R}((1 +
\xi^j)e^{-\gamma\xi)}} \leq C(1 + \tau ^{k-j}),
\end{equation}
if $1 \leq i \leq r - 1$.  The nonhomogeneous term in (\ref{E5.8a}) has
the same form as in (\ref{E5.9}), but the norm has to be replaced by the
norm in $E^1_{\Bbb R^+}((1 + \xi^j)e^{-\gamma\xi})$.

We need similar decompositions of $\{\tilde u_j\}^\infty_{j=0}$ and
$\{\Delta
u_j(\xi,0)\}^\infty_{j=0}$.  The expansion of $\sum \epsilon
^ju^{RS,i+1}_j$
into $\sum \epsilon ^j\tilde u_j$ can be divided into two steps.  Let
\begin{align*}
\sum ^\infty_{j=0} \epsilon ^j\tilde {\tilde u}_j(\xi,\tau) &= \sum
^\infty_{j=0} \epsilon ^ju^{RS,i+1}_j(\epsilon \xi +
\stackrel{*}{\eta}\!{}^i_0,\tau),\\
\sum ^\infty_{j=0} \epsilon ^j\tilde u_j(\xi,\tau) &= \sum ^\infty_{j=0}
\epsilon ^j\tilde {\tilde u}_j(\xi + \sum ^\infty_{\ell = 0}  \epsilon
^\ell \stackrel{*}{\eta}\!{}^i_{\ell +1} (\tau),\tau).
\end{align*}
We then have
\begin{equation}\label{E5.10}
\tilde u_k(\xi,\tau) = \sum _{\delta,\alpha} C_{\alpha \delta}
D^{|\alpha|}_x \tilde {\tilde u}_\delta (\xi +
\stackrel{*}{\eta}\!{}^i_1(\tau))[\stackrel{*}{\eta}]^\alpha
\stackrel{\rm def}{=} \sum ^k_{\delta = 0} \tilde u_{k\delta}.
\end{equation}
Here $\delta \geq 0$ is an integer, $[\stackrel{*}{\eta}]^\alpha =
[\stackrel{*}{\eta}\!{}^i_2(\tau)]^{\alpha _1
} \ldots [\stackrel{*}{\eta}\!{}^i_{k+1}(\tau)]^{\alpha_k}$, $C_{\alpha
\delta}$
is a constant and $\delta + \displaystyle \sum ^k_{j=1} j \alpha _j = k$.
$\tilde u_{k\delta}$ consists of derivatives of $\tilde {\tilde u}_\delta$
only
and  \begin{equation}\label{E5.11}
| \tilde u_{k\delta}|_{E^m((1+|\xi|^\delta))} \leq C_{mk\delta}(1 + \tau
^{k-\delta}),\ m \geq 0.
\end{equation}
In comparison, $\sum \epsilon ^j\overline u_j(\xi)$ is also defined in a
similar way
$$
\sum ^\infty_{j=0} \epsilon ^j \overline u_j(\xi) = \sum ^\infty_{j=0}
\epsilon
^j \overline {\overline u}_j(\xi + \sum ^\infty_{\ell = 0} \epsilon ^\ell
\overline \eta ^i_{\ell +1}).
$$
Exactly like (\ref{E5.10}), we have

\begin{eqnarray}
\overline u_k(\xi)  &  = &  \sum _{\delta , \alpha} C_{\alpha \delta}
D^{|\alpha|}_x \overline {\overline u}_\delta  (\xi +  \overline \eta
^i_1)[\overline \eta ]^\alpha \stackrel{\rm def}{=} \sum ^k_{\delta = 0}
\overline u_{k\delta}\\
&& |\overline u_{k\delta}|_{E^m(1+|\xi|^\delta)} \leq \overline
C_{mk\delta}.\nonumber
\end{eqnarray}
Recall that $\stackrel{*}{\eta}\!{}^i_j(0)  = \overline \eta ^i_j$ and
$|\overline {\overline u}_\delta (\xi) - \tilde {\tilde u}_\delta
(\xi,0)|_{E^m((1 + |\xi|^\delta)e^{-\gamma \xi})} \leq C$, cf. H8).
We have
\begin{equation}\label{E5.13}
\Delta  u_k(\xi,0) = \sum ^k_{j=0} \Delta u_{kj} (\xi,0) \stackrel{\rm
def}{=}
\sum ^k_{j=0}  [\overline u_{kj}(\xi)  - \tilde u_{kj}(\xi,0)],
\end{equation}
with
\begin{equation}\label{E5.14}
|\Delta u_{kj}(\cdot , 0)|_{E^m((1+|\xi|^j)e^{-\gamma \xi})} \leq C.
\end{equation}
For $1\leq i\leq r-1$ we solve
\begin{equation}\label{E5.15}
\bg{aligned}
& \Delta u_{kj\tau}  =  \Delta u_{kj\xi\xi} + V^i(\overline \eta ^i) \Delta
u_{kj\xi} + f_{0u}(\tilde u_0,\overline \eta ^i)\Delta u_{kj} +
G_{kj}\\
& \Delta u_{kj}(\xi,0)  =  \overline u_{kj}(\xi) - \tilde u_{kj}(\xi,0).
\end{aligned}
\end{equation}
For $i=0$, we solve
\begin{equation}\label{E5.15a}
\bg{aligned}
& \Delta u_{kj\tau}  =  \Delta u_{kj\xi\xi} + V^i(\overline \eta
^i) \Delta
u_{kj\xi} + f_{0u}(u_0^{SSi},\overline \eta ^i)\Delta u_{kj} +
G_{kj},\\
& \Delta u_{kj}(\xi,0)  =  \overline u_{kj}(\xi) - \tilde
u_{kj}(\xi,0),
\end{aligned}
\end{equation}
\bg{equation*}\tag{\ref{E5.15a}${}_1$}
\Delta u_{kj\xi}(0,\tau) = -\tilde u_{kj\xi}(0,\tau).
\end{equation*}

Let $1 \leq i \leq r - 1$ first.  Since $\tilde u _0 \to p^{i+1}(\overline
\eta
^i)$ as $\tau \to \infty$, equation (\ref{E5.15}) is exponentially stable in
the
space $E^2_{\Bbb R}((1 + |\xi|^j)e^{-\gamma\xi})$.  Therefore, from
\lemref{LB.3a}, with $\theta=1,\:\beta=\frac{1}{2}$,
$$
|\Delta u_{kj}|_{E^2_{\Bbb R} ((1 + |\xi|^j)e^{-\gamma\xi})} \leq C(1 +
\tau ^{k-j}).
$$
Thus (\ref{E5.7}) has been proved for $1 \leq i \leq r - 1$.

Let $i = 0$.  We consider (\ref{E5.15a}) in $\xi \geq 0$.  Estimate in
(\ref{E5.9}) is replaced by
$$
|G_{kj}|_{E^1_{\Bbb R^+}((1+\xi^j)e^{-\gamma\xi})} \leq C(1 + \tau
^{k-j}).
$$
To solve (\ref{E5.15a}) and (\ref{E5.15a}${}_1$),  let
$\tilde u_{kj\xi}(0,\tau) = \varphi _j(\tau)$. We have $|D^\alpha \varphi
_j(\tau)|\newline \leq C(1 + \tau ^{k-j})$, $\alpha=0,\,  1$.
Let $\Phi (\xi,\tau)$ be the solution to the elliptic system
with a boundary condition at $\xi=0$,
$$
\begin{array}{l}
\Phi_{\xi\xi} + V^i(\overline \eta ^i)\Phi _\xi + f_{0u}(q^i(\xi,\overline
\eta ^i), \overline \eta ^i)\Phi = 0,\\
\Phi _\xi (0,\tau) = - \varphi _j(\tau)
\end{array}
$$
Then from \lemref{LB.2}.(ii), there exists a unique solution $\Phi$ such that
$$
  |\partial_\tau^\alpha \Phi|_{E^2_{\Bbb R^+} ((1 + \xi^j)e^{-\gamma \xi})}
\leq C(1 + \tau ^{k-j}), \quad   \alpha =0,\,1.
$$
The solution $\Delta u_{kj} = \Phi + \Psi$, with
$$
\begin{array}{rcl}
\Psi_\tau & = & \Psi_{\xi\xi} + V^i(\overline \eta ^i) \Psi_\xi +
f_{0u}(u_0^{SSi},\overline \eta ^i)\Psi + G_{kj} - \Phi_\tau\\
& + &  [f_{0u}(u_0^{SSi},\overline \eta ^i)
-f_{0u}(q^i(\xi,\overline \eta ^i),\overline \eta ^i) ]\Phi\\
\Psi _\xi(0,\tau)  & = &  0,\\
\Psi (\xi, 0) & = & \Delta u_{kj}(\xi,0) - \Phi(\xi,0).
\end{array}
$$
Since the linear equation for $\Psi$ is asymptotically stable, and the
forcing
terms are in $E^1_{{\Bbb R^+}
((1 + |\xi|^j)e^{-\gamma\xi})}$ with norms bounded by $C(1 + \tau ^{k-j})$,
from \lemref{LB.3a} again,
$$
|\Psi|_{E^2_{\Bbb R^+}((1+\xi^j)e^{-\gamma\xi})} \leq C(1 + \tau ^{k-j}).
$$
This proves the case $i = 0$.  The matching of $(SS)^i$ with $(RS)^{i+1}, 0
\leq i \leq r - 1$ has been proved by induction.  The matching of $(SS)^i$
with
$(RS)^i$, $1 \leq i \leq r$ can be treated similarly.

\bg{thm}\label{T5.4}
Let $\Delta u^i_j=u^{SSi}_j-\tilde u^{RSi}_{j,1},\:1\leq i\leq r$ or
$u^{SSi}_j-\tilde u^{RSi}_{j,2},0\leq i\leq r-1$,
where $\tilde u^{RSi}_{j,1}$ and
$\tilde u^{RSi}_{j,2}$ are defined in (\ref{E5.4a}). Then there exists
$\gamma >0$ such that for all integers $\alpha\leq 2,\,\beta\leq 1$,
$$
 \vert\partial^\alpha_\xi\Delta u^i_j\vert + \vert\partial^\beta_\tau \Delta
 u^i_j\vert \leq C_{j} (1+|\xi|^j+\tau^j)e^{-\gamma |\xi|}.
$$
\end{thm}

\subsection{Constructing a Pseudo Solution}
A pseudo solution is a piecewise smooth function that almost satisfies
(\ref{E1.1}) with small residual error
$$\epsilon u_t-\epsilon^2 u_{xx}-f(u,x,\epsilon),
$$
in the interior of each subregion where the function is smooth, and small
jump error at their common boundaries. By truncating the formal series, we
can construct pseudo solutions with arbitrary accuracy.

Let
\bg{align*}
u^{RRi,m}(x,\epsilon)=\sum_{j=0}^m \epsilon^j u_j^{RRi}(x),
&\quad u^{SRi,m}(\xi,t,\epsilon)=\sum_{j=0}^m\epsilon^j u_j^{SRi}(\xi,t),\\
u^{RSi,m}(x,\tau,\epsilon)=\sum_{j=0}^m\epsilon^j u_j^{RSi}(x,\tau),
&\quad u^{SSi.m}(\xi,\tau,\epsilon)=\sum_{j=0}^m\epsilon^j
u_j^{SSi}(\xi,\tau),\\
\eta^{i,m}(t,\epsilon)=\sum_{j=0}^m\epsilon^j\eta_j^i(t),
&\quad \stackrel{*}{\eta}\!{}^{i,m}(\tau,\epsilon)=
\sum_{j=0}^m\epsilon^j \stackrel{*}{\eta}\!{}^i_j(\tau).
\end{align*}

We use $\eta^{i,m}$ and $\stackrel{*}{\eta}\!{}^{i,m}$ as abbreviations for
$\eta^{i,m}(t,\epsilon)$ and $\stackrel{*}{\eta}\!{}^{i,m}(\tau,\epsilon)$.
Let $0<\beta<1$ be a constant. Let the width of the initial and
internal/boundary layers be $O(\epsilon^\beta)$.
Define the subregions
\bg{align*}
& (RR)^{i,m}=\{t>\epsilon^\beta,\quad x\in(\eta^{i-1,m}+\epsilon^\beta,\,
\eta^{i,m}-\epsilon^\beta)\},\quad 1\leq i\leq r,\\
& (SR)^{i,m}=\{t>\epsilon,\quad x\in(\eta^{i,m}-\epsilon^\beta,\,
\eta^{i,m}+\epsilon^\beta)\cup [a,\,b]\},\quad 0\leq i\leq r,\\
& (RS)^{i,m}=\{0\leq t<\epsilon^\beta,\quad x\in(\stackrel{*}{\eta}\!{}
^{i-1,m+1}+\epsilon^\beta,\,\stackrel{*}{\eta}\!{}^{i,m+1}-\epsilon^\beta)\},
\quad 1\leq i\leq r,\\
& (SS)^{i,m}=\{0\leq t<\epsilon^\beta,\quad
x\in(\stackrel{*}{\eta}\!{}^{i,m+1}-\epsilon^\beta,\,
\stackrel{*}{\eta}\!{}^{i,m+1}+\epsilon^\beta)\},\quad 0\leq i\leq r.
\end{align*}

\bg{thm}\label{T5.5}
For all $m\geq 0$, let
$$U^m(x,t,\epsilon)=\bg{cases}
u^{RRi,m}(x,t,\epsilon), &\text{if }(x,t)\in (RR)^{i,m},\\
\dis u^{SRi,m}(\frac{x-\eta^{i,m}}{\epsilon},t,\epsilon),
&\text{if }(x,t)\in (SR)^{i,m},\\
\dis u^{RSi,m}(x,\frac{t}{\epsilon},\epsilon), &\text{if }(x,t)\in (RS)^{i
,m},\\
\dis u^{SSi,m}(\frac{x-\stackrel{*}{\eta}{}^{i,m+1}}{\epsilon},
\frac{t}{\epsilon},\epsilon), &\text{if }(x,t)\in (SS)^{i ,m}.
\end{cases}
$$
Then $U^m$ is a pseudo solution to (\ref{E1.1}) with
residual and jump errors $O(\epsilon^{\beta(m+1)})$.
\end{thm}

\bg{pf} The residual errors can be evaluated by substituting $U^m$ into
(\ref{E1.1}) and expanding in powers of $\epsilon$. For example, in
$(SS)^{i,m}$, the error is bounded by
$$
 C\,\epsilon^{m+1}\,\sup\{ 1+|\xi|^{m+1}+|\tau|^{m+1}\; :\;
|\xi|<\epsilon^{\beta-1},\,|\tau|<\epsilon^{\beta-1} \} \leq
C \,\epsilon^{\beta(m+1)}.
$$

The estimates of jump errors use the matching of adjacent layers. For
example, the jump between $(SS)^{i,m}$ and $(RS)^{i+1,m}$ at the boundary
\bg{align*}
 \Gamma & =\{x=\stackrel{*}{\eta}\!{}^{i,m+1}+\epsilon^\beta,\, 0\leq \tau\leq
\epsilon^{\beta-1}\}\\
&=\{\xi=\epsilon^{\beta-1},\,0\leq \tau\leq \epsilon^{\beta-1}\},
\end{align*}
where $\xi=\frac{x-\stackrel{*}{\eta}{}^{i,m+1}}{\epsilon}$, is
\bg{equation}\label{E5.16}
\bg{split}
&\vert \sum_0^m(\epsilon^j(u^{SSi}_j(\xi,\tau)-u^{RS,i+1}_j(x,\tau))\vert\\
\leq &\vert\sum_0^m\epsilon^j(u^{SSi}_j(\xi,\tau)-
   \tilde u^{RSi}_{j,2}(\xi,\tau))\vert +\vert \sum_0^m\epsilon^j(\tilde
u^{RSi}_{j,2}(\xi,\tau)-u^{RS,i+1}_j(x,\tau))\vert.
\end{split}
\end{equation}
The first term of the above is bounded by
$$
 O(\sum_0^m \epsilon^j\,\sup\{(1+|\xi|^j+|\tau|^j)e^{-\gamma\xi}\})
=O(e^{-\gamma\epsilon^{\beta-1}})
$$
where $\gamma>0$ is a constant, due to the matching condition
\thmref{T5.4}. For the second term, observe that
\bg{align*}
\sum_0^m\epsilon^j u^{RS,m+1}_j (\epsilon\xi+\stackrel{*}{\eta}\!{}^{i,m+1},
\tau) &= \sum_0^m\epsilon^j u^{RS,m+1}_j
(\epsilon\xi+\stackrel{*}{\eta}\!{}^{i},\tau) +
O(\epsilon^{\beta(m+1)})\\
&=\sum_0^m\epsilon^j \tilde u^{RSi}_{j,2}(\xi,\tau)
+O(\epsilon^{\beta(m+1)}).
\end{align*}
Here the $O(\epsilon^{\beta(m+1)})$ terms are caused by truncating
$\stackrel{*}{\eta}\!{}^{i}$. This gives the desired estimate on the
jump error. The other jump errors can be estimated similarly.
\end{pf}

Finally we mention that using composite expansion technique we can
construct pseudo solutions with only residual error but no jump error
in the entire region $t\geq 0,\:x\in [a,\,b]$. See \cite{lin89} and
\cite{eckhaus77,eckhaus79} for more details.

\endinput


\section{Proof of the lemmas}

\bg{pf*}{Proof of \lemref{LB.1}}  Consider a system in ${\Bbb R}^{2n}$
that is  equivalent to (\ref{EB.6}):
\begin{equation}\label{E6.1}
\begin{array}{lll}
u_\xi & = & v,\\
v_\xi & = & -Df(p)u - Vv.
\end{array}
\end{equation}
Let $\lambda$ be an eigenvalue for $J = \left
(\begin{array}{cc}
0 & I \\ -Df(p) & -V\end{array}\right )$.
Then, $\det(\lambda I- J)= \displaystyle \prod ^n_{i=1}(\lambda ^2 +
V\lambda + \mu_i)$,
where $\mu_i,\, 1 \leq i \leq n$ are the eigenvalues for $Df(p)$.  Since
$Re\mu_i
\leq - \sigma _0$, it is elementary to show that $\lambda ^2 + V\lambda +
\mu_i = 0$ has two roots $\{\lambda _{i1}, \lambda _{i2}\}$ with $Re \lambda
_{i1} \leq - V - \sqrt{V^2+4\sigma _0} < \sqrt{V^2 + 4\sigma  _0 } - V \leq
Re\lambda _{i2}$.  Therefore (\ref{E6.1}) has exponential dichotomy with
stable and unstable spaces both $n$-dimensional, so does (\ref{EB.6}).

Since  $Df(q(\xi)) \to Df(p^i),\, i = 1$ or $2$ as $\xi \to - \infty$ or
$+\infty$, by a perturbation theory of exponential dichotomy, cf.
\cite{palmer84},  (\ref{EB.7})
has exponential dichotomies on $R^-$ and ${\Bbb R}^+$ respectively.  The
stable
and unstable spaces, ${\cal R}P_s(t)$ and ${\cal R}P_u(t)$ are
$n$-dimensional. The rate of decay on ${\cal R}P_s(t)$ and ${\cal
R}P_u(t)$ can be any constant $0<\alpha_1<\alpha$. Since $\alpha$ can be
arbitrary close to $\sqrt{V^2+4\sigma_0}-|V|$, so is $\alpha_1$.
\end{pf*}

We now consider the following systems
\begin{equation}\label{E6.2}
\begin{array}{lll}
u_\xi & = & v,\\
u_\xi & = & -Vv - Df(p) u + g,
\end{array}
\end{equation}
\begin{equation}\label{E6.4}
\begin{array}{lll}
u_\xi & = & v\\
v_\xi & = & -Vv - Df(q(\xi))u + g
\end{array}
\end{equation}
Let $T(\xi,s)$ be the solution operator for (\ref{E6.4}).

\bg{pf*}{Proof of \lemref{LB.2}}    (i) Let $g\in X = E^m_{\Bbb R}(w)$.
Let $L_{p}u = u_{\xi\xi}+V u_\xi +Df(p)u$, with $D(L_p)= E^{m+2}_{\Bbb
R}(w)$.  From \lemref{LB.1}, system (\ref{E6.2})
that is equivalent to $L_p u=g$ has an exponential dichotomy on ${\Bbb R}$.
Let $P_u$ and $P_s$ be the
projection to the unstable and stable spaces.  Let
\begin{equation}\label{E6.3}
\left ( \begin{array}{c}
u\\ v\end{array}\right ) (\xi) = \int^\xi_{-\infty} e^{J(\xi-s)} P_s \left
(
\begin{array}{c}
0\\ g(s)\end{array} \right ) ds + \int^\xi _{\infty} e^{J(\xi-s)} P_u \left
(
\begin{array}{l}
0 \\ g(s)\end{array} \right ) ds.
\end{equation}
Using the exponential estimates on $\|e^{J(\xi - s)^\infty} P_s\|$
and $\|e^{J(\xi-s)}P_u\|$, and \lemref{LB.0}, we can verify that $(u,v)\in
E^{m+2}_{\Bbb
R}(w)\times E^{m+1}_{\Bbb R}(w)$ and  $(u,v)$ solves (\ref{E6.2}).
Details will be omitted.  Therefore ${\cal
R}(L_p) = X$.  On the other hand, if $u$ is a solution to $L_{p}u = g$, then
$(u,u_\xi) \in
E^{m+2}_{\Bbb R}(w) \times E^{m+1}_{\Bbb R}(w)$ is a solution to
(\ref{E6.2}).
It is standard to show $(u,u_\xi)$ is given by (\ref{E6.3}).
This proves that $Ker(L_p) = \{0\}$.

\noindent (ii) Let $g \in X = E^m_{\Bbb R^+}(w)$.  All the solutions of
(\ref{E6.4}) with $u \in E^{m+2}_{\Bbb R^+}(w)$ are given by
\bg{align*}
\left ( \begin{array}{c}
u\\ v\end{array} \right ) (\xi) &= T(\xi,0) \left ( \begin{array}{c}
u_0\\ v_0\end{array} \right ) + \int^\xi _0 T(\xi,s)P_s\left
(\begin{array}{l}
0\\ g(s)\end{array} \right ) ds\\ &+ \int^\xi _\infty T(\xi,s)P_u\left (
\begin{array}{c}
0\\ g(s)\end{array} \right )ds,
\end{align*}
where $(u_0,v_0) \in {\cal R}P_s(0)$. Let $\Pi(u,v)=v$ be the projection
from ${\Bbb R}^{2n}\sim {\Bbb R}^n\times{\Bbb R}^n$ to ${\Bbb R}^n$.
 We can show that  $\Pi:{\cal R}P_s \to {\Bbb R}^n$ is a homeomorphism. If
not, then there exists a nontrivial $\dis \bg{pmatrix} u_0\\v_0 \end{pmatrix}
\in {\cal R}P_s(0)$, such that $v_0=0$. Thus, there exists a nontrivial bounded
solution to (\ref{EB.7}) with $u_\xi(0)=0$. This is a contradiction.
Based on what have been proved, there exists a unique
$\left ( \begin{array}{c} u_0 \\ v_0\end{array} \right ) \in {\cal R}P_s$
such that
$$
\Pi \left ( \begin{array}{c}
u_0 \\ v_0\end{array} \right ) = v_0 = \phi - \Pi \int^0_\infty
T(0,s) P_u\left ( \begin{array}{c} 0 \\ g(s)\end{array} \right ) ds.
$$
The desired solution can be obtained by such $\left ( \begin{array}{c} u_0
\\ v_0\end{array} \right )$.

The case $X = E^m_{\Bbb R^-}(w)$ can be treated similarly.

Finally the estimates on $\|u\|_{E^{m+2}(w)}$ in both cases (i) and (ii)
come
from Banach's closed graph theorem.
\end{pf*}

\bg{pf*}{Proof of \lemref{LB.3}} Consider (\ref{E6.4}) but with $\xi\in
{\Bbb R}$, $g\in X= E^m_{\Bbb R}(w)$. Equation $L_{q}u=g$ is then
equivalent to (\ref{E6.4}). From \lemref{LB.1},
(\ref{E6.4}) has exponential dichotomies on ${\Bbb R^+}$ and ${\Bbb R^-}$.
Also if $g=0$, $(q_\xi,q_{\xi\xi})$ is the only bounded solution to
(\ref{E6.4}),
up to constant multiples.  It follows from the same argument as in
\cite{palmer84}, which treats the case $w \equiv 1$, that (\ref{E6.4}) has a
solution $(u,v) \in E^{m+2}_{\Bbb R}(w) \times E^{m+1}_{\Bbb R}(w)$ if and
only
if $\langle \left ( \begin{array}{c} 0 \\ g \end{array} \right ), \left (
\begin{array}{c} \psi _1 \\ \psi_2\end{array} \right ) \rangle  _{L^2} =
0$,
where $(\psi _1, \psi_2)$ is a unique (up to constant multiples) bounded
solution to the adjoint equation of (\ref{E6.4})
\begin{equation}\label{E6.4a}
\begin{aligned}
\psi_{1\xi} & =  Df^\tau(q(\xi))\psi_2,\\
\psi_{2\xi} & =  -\psi_1 + V\psi_2.
\end{aligned}
\end{equation}
It is now clear that $\langle g,\psi_2 \rangle  = 0 \Leftrightarrow g \in
{\cal R}(L_q)$. The equation for $\psi_2$ is
$\psi_{2\xi\xi} - V\psi _{2\xi} + Df^\tau(q(\xi)) \psi_2 = 0$.
\end{pf*}

\bg{pf*}{Proof of \lemref{LB.3a}} Using the definition of the exponential
stability right before \lemref{LB.3a}, we have
\bg{align*}
\|T(t,0)u_0\|_\theta & \leq Ke^{-\alpha t}\|u_0\|_\theta \leq C,\\
\|\int_0^t T(t,s)g(s)\,ds\|_\theta &\leq
\int_0^t\| T(t,\frac{t+s}{2}) T(\frac{t+s}{2},s)g(s)\|_\theta\,ds\\
&\leq \int_0^t Ke^{-\alpha(t-s)/2}K(1+(\frac{t-s}{2})^{\beta-\theta})\cdot
C(1+s^k)\,ds\\
&\leq C(1+t^k).
\end{align*}
The desired estimate of $u(t)$ follows from the variation of constant
formula.
\end{pf*}

\bg{pf*}{Proof of \lemref{LB.4}}
Fro any $\gamma \in {\Bbb R}$, the locus of ${\cal P} = \{\lambda ^2 \vert
Re \lambda = 2|\gamma|+1\}$ is a parabola. Let $\Sigma =
\{|arg (\lambda - \sigma _1)| < {\pi \over 2} + \delta \},\: 0 < \delta <
{\pi \over 2}$, be a sector. Let $\sigma_1>0$ be sufficiently large
 such that $\Sigma \cap{\cal P} = \emptyset$.  Then
$\lambda \in \Sigma $ implies that $Re \sqrt{\lambda} > 2 |\gamma|+1$,
where $\sqrt{\lambda}$ is in the branch with $|\arg \sqrt{\lambda}|<\pi/2$.

\noindent  (i)  Let $g \in X = E_{\Bbb R}(w),\, w=(1+|\xi|^j)e^{-\gamma\xi},\,
\lambda \in \Sigma$. Consider $u_{\xi\xi} - \lambda u = g$, and its
equivalent system
\begin{equation}\label{E6.5}
\begin{array}{lll}
u_\xi & = & \sqrt{\lambda} v,\\
v_\xi & = & \sqrt{\lambda} u + g/\sqrt{\lambda}.
\end{array}
\end{equation}
The eigenvalues for $H = \left ( \begin{array}{cc}
0 & \sqrt{\lambda} I\\
\sqrt{\lambda} I & 0 \end{array} \right )$ are $\mu = \pm \sqrt{\lambda}$,
each is of multiplicity $n$.  Also $|Re\mu| > 2 |\gamma|+1$.  Therefore
(\ref{E6.5}) has an exponential dichotomy on ${\Bbb R}$ , with $n$-dimensional
stable and unstable subspaces.

For the matrix $H$, ${\cal R}P_u = \left \{\left ( \begin{array}{c} w \\
w\end{array} \right ) | w \in {\Bbb R}^n \right \}$ that is associated to
eigenvalue $\sqrt {\lambda}$ and ${\cal R}P_s = \left \{\left (
\begin{array}{c}
w \\ -w\end{array} \right ) | w \in {\Bbb R}^n \right \}$ that is
associated to
eigenvalue $-\sqrt{\lambda}$.  Thus, $P_s \left ( \begin{array}{c} u \\ v
\end{array} \right ) = \left ( \begin{array}{c} (u - v)/2\\
(v-u)/2\end{array}\right )$ and $P_u\left ( \begin{array}{c} u \\
v\end{array}\right ) = \left ( \begin{array}{c}
(u + v)/2 \\ (u + v)/2\end{array} \right )$ with $|P_u| + |P_s| \leq C$.
The constant $C$ is independent of $\lambda \in \Sigma$.  Thus, in the
following, the constant $K$ is independent of $\lambda \in \Sigma$.
\begin{equation}\label{E6.6}
\begin{array}{lr}
|e^{H\xi}P_s| \leq Ke^{-Re\sqrt{\lambda}\, \xi}, & \xi \geq 0,\\
|e^{H\xi}P_u| \leq Ke^{Re\sqrt{\lambda}\, \xi}, & \xi \leq 0.
\end{array}
\end{equation}
Using (\ref{E6.6}), we can show that the integrals in (\ref{E6.7})
converge and define a solution to (\ref{E6.5}).
\begin{equation}\label{E6.7}
\left ( \begin{array}{c} u \\ v \end{array} \right ) (\xi)  =
\int^\xi_{-\infty} e^{H(\xi-s)} P_s \left ( \begin{array}{c} 0 \\
g(s)/\sqrt{\lambda} \end{array} \right ) ds + \int^\xi_\infty
e^{H(\xi-s)}P_u
\left (\begin{array}{l} 0 \\ g(s)/\sqrt{\lambda}\end{array} \right ) ds.
\end{equation}
In fact, using the estimates (\ref{E6.6}) and \lemref{LB.0}, we have
$$
\begin{array}{lll}
\left |\left ( \begin{array}{c} u \\ v \end{array}\right ) (\xi)\right | &
\leq &
\int^\xi_{-\infty} Ke^{-Re\sqrt{\lambda}(\xi-s)}\|g\|_{E(w)}\cdot  {1\over
|\sqrt{\lambda }|} (1 + |s|^j) e^{-\gamma s}ds\\
& + & \int^\infty _\xi Ke^{-Re\sqrt{\lambda}(s-\xi)}
\|g\|_{E(w)}{1\over|\sqrt{\lambda}|}(1+|s|^j)e^{-\gamma s}ds\\
& \leq & {K_1\|g\|_{E(w)} \over (Re\sqrt{\lambda}-|\gamma|)^{j+1}\cdot
|\sqrt{\lambda}|} \cdot  (1 + |\xi|^j)e^{-\gamma\xi}.
\end{array}
$$
Here $K_1$ is a constant that depends on $j$. From
$Re\sqrt{\lambda}>2|\gamma| +1$, we have $Re\sqrt{\lambda}-|\gamma|>1$ and
$Re\sqrt{\lambda}-|\gamma|>\frac{1}{2}Re\sqrt{\lambda}$. Therefore the above
is bounded by
$$
  {2K_1\|g\|_{E(w)}\over Re\sqrt{\lambda} \cdot | \sqrt{\lambda}|}\cdot
  (1 + |\xi|^j)e^{-\gamma \xi}.
$$
Since $\sigma_1 >0$, $ |arg \lambda| \leq |arg(\lambda -\sigma _1)| \leq
{\pi \over 2} + \delta$, and $|arg\sqrt{\lambda}| \leq {\pi \over 4} + {\delta
\over 2} < {\pi \over 2}$.  Thus, $Re\sqrt{\lambda} > |\sqrt{\lambda}| \cos
({\pi \over 4} + {\delta \over 2}) = C_0|\sqrt{\lambda}|$.  From this
$$
\left | \left ( \begin{array}{c} u \\ v \end{array}\right ) (\xi)  \right |
\leq
{2K_1\|g\|_{E(w)} \over C_0|\lambda|} (1 + |\xi|^j)e^{-\gamma \xi}.
$$
Thus $\|u\|_{E(w)} \leq {c \over |\lambda|} \|g\|_{E(w)} \leq {c \over
|\lambda-\sigma_1|} \|g\|_{E(w)}$.  The solution in (\ref{E6.7}) is also
unique in $E_{\Bbb R}(w)$ since (\ref{E6.5}) has an exponential dichotomy
on
$\Bbb R$.  This proves the Lemma when $X = E_{\Bbb R}(w)$.

\noindent (ii) Let $g \in X = E_{\Bbb R^+}(w), \lambda \in \Sigma$.  Consider
$u_{\xi\xi} - \lambda u = g$ with the boundary condition $u_\xi(0) = 0$.
If
$\left ( \begin{array}{c} u\\ v\end{array} \right ) \in E^m_{\Bbb R^+}(w)$
is a
solution for (\ref{E6.5}), it can be expressed as
\bg{align*}
\left ( \begin{array}{c} u \\ v \end{array} \right ) (\xi) &= e^{H\xi} \left
(
\begin{array}{c} u_0\\ v_0\end{array} \right ) + \int^\xi_0 e^{H(\xi-s)}
P_s\left ( \begin{array}{c} 0 \\ g(s)/\sqrt{\lambda} \end{array} \right )ds\\
& + \int^\xi _\infty e^{H(\xi-s)} P_u \left ( \begin{array}{c} 0 \\ g(s)/\sqrt
{\lambda} \end{array} \right ) ds.
\end{align*}
Here $\left ( \begin{array}{c} u_0 \\ v_0 \end{array} \right ) \in
{\cal R}P_s$ has to be determined so that $v(0) = 0$.  To this end, choose
$$
v_0  = - \Pi \int^0_\infty
e^{-Hs} P_u\left ( \begin{array}{c} 0 \\ g(s)/\sqrt{\lambda} \end{array}
\right) ds
$$
and $u_0 = - v_0$, then $\left ( \begin{array}{c} u_0 \\ v_0
\end{array}
\right ) \in {\cal R}P_s$.  Similar to (i), $\left |\left (
\begin{array}{c} u_0
\\ v_0\end{array}\right ) \right |_{{\Bbb R}^{2n}} \leq  C\|g\|_{E_{\Bbb
R}^+(w)}/|\lambda|$.  With this $\left ( \begin{array}{c} u_0\\
v_0\end{array} \right )$, we have
$$
\left |\left ( \begin{array}{c} u\\ v\end{array} \right )\right  |_{E_{\Bbb
R^+}(w)} \leq C\|g\|_{E(w)/|\lambda|} \leq C\|g\|_{E_{\Bbb R^+}(w)}/|\lambda -
\sigma _1|. $$
This proves the case $X = E_{\Bbb R^+}(w)$.  The same argument can be used
to
the case $X = E_{\Bbb R^-}(w)$.

\noindent (iii) Let $g \in X = B_{\Bbb R}(w), \lambda \in \Sigma$.  If $\left
(
\begin{array}{c} u \\ v \end{array} \right ) \in B_{\Bbb R}(w)$ is a
solution
to (\ref{E6.5}), it is expressed by (\ref{E6.7}).  Let $g = g(\infty) +
g_1(\xi)$, where $g(\infty) = \displaystyle \lim_{\xi \to \infty} g(\xi)$.
Then
$g_1 \in E_{\Bbb R}(w)$.  It follows that $\left ( \begin{array}{c} u\\
v\end{array} \right ) = \left ( \begin{array}{c} u_1 \\ v_1\end{array}
\right )
+ \left ( \begin{array}{c} u_2 \\ v_2\end{array} \right )$, where $\left (
\begin{array}{c} u_2 \\ v_2 \end{array} \right ) = \left ( \begin{array}{c}
-g(\infty)/\lambda \\ 0\end{array} \right )$ is a constant solution and
$$
\left ( \begin {array}{c} u_1 \\ v_1 \end{array} \right ) (\xi) = \int^\xi
_{-\infty} e^{H(\xi-s)} P_s \left ( \begin{array}{c} 0 \\
g_1(s)/\sqrt{\lambda}\end{array} \right ) ds + \int^\xi_\infty e^{H(\xi-s)}
P_u
\left ( \begin{array}{c} 0 \\ g_1(s)/\sqrt{\lambda} \end{array} \right )
ds.
$$
{}From (i), $\|u_1\|_{E_{\Bbb R}(w)} \leq C\|g_1\|_{E_{\Bbb R}(w)
/|\lambda|}$.
Also, $|u_2| \leq |g(\infty)|/|\lambda|$.
Thus, $\|u\|_{B_{\Bbb R}(w)} \leq C\|g\|_{B_{\Bbb R}(w)}/|\lambda| \leq
C\|g\|_{B_{\Bbb R}}(w)/|\lambda - \sigma  _1|$.  This proves the case $X =
B_{\Bbb R}(w)$.  The cases $X = B_{\Bbb R^\pm}(w)
$ can be handled similar to (ii).
\end{pf*}   

\bg{pf*}{Proof of \lemref{LB.4a}} (i) Let $w=1+|\xi|^j$,
$X = E_{\Bbb R}(w)$ and $D_A =
E^2_{\Bbb R}(w)$ first.  Let $u_0 \in E^1_{\Bbb R}(w), v = D_\xi u_0 \in
E_{\Bbb R}(w)$.  Let
$$
I(t) = t^{1/2} Ae^{At} u_0 = t^{1/2}D^2_\xi e^{At}u_0 = t^{1/2} D_\xi
e^{At}v.
$$
Using the fundamental solution for the heat equation
$$
\begin{array}{lll}
I(\xi,t) & = & t^{1/2} D_\xi \int^\infty_{-\infty} (2\sqrt{\pi t})^{-1}
\exp (-{x^2\over 4t})v(\xi - x)dx\\
& = & (2\sqrt{\pi})^{-1} \int^\infty_{-\infty}\exp (-{x^2\over 4t})v(\xi -
x)(-{2x \over 4t})dx.
\end{array}
$$
Let ${x^2\over 4t} = \eta$, $x = \pm\sqrt{4t\eta}$.
$$
\begin{array}{rcl}
|2\sqrt{\pi} I(\xi,t)| & \leq & | - \int^0_\infty e^ {-\eta }v(\xi +
\sqrt
{4t\eta})d\eta\\
&& \mbox {} - \int^\infty_{0} e^ {-\eta} v(\xi - \sqrt{4t\eta})d\eta
|\\
& = & |\int^\infty_{0} e^ {-\eta} [v(\xi + \sqrt{4t\eta}) -
v(\xi)]d\eta\\
&& \mbox {} - \int^\infty_{0} e^ {-\eta} [v(\xi - \sqrt{4t\eta}) -
v(\xi)]d\eta|\\
|2\sqrt{\pi}(1 + |\xi|^j)^{-1}I(\xi,t)| & \leq & |\int^M_0 + \int^\infty_M
e^ {-\eta } | v(\xi + \sqrt{4t\eta}) - v(\xi) | (1 + |\xi|^j)^{-1}d\eta| \\

&+&  |\int^M_0 + \int^\infty_M e^ {-\eta}| v(\xi - \sqrt{4t\eta})
- v(\xi) | ( 1+ |\xi|^j)^{-1}d\eta |\\
&=&  I_1 (\xi, t) + I_2(\xi,t).
\end{array}
$$
We show $\displaystyle \lim_{t\to 0^+} \displaystyle \sup_{\xi} I_1(\xi,t)
= 0$.  Similar arguments will show $\displaystyle \lim _{t \to 0^+}
\displaystyle
\sup_{\xi} I_2(\xi,t) = 0$, thus $\displaystyle \lim_{t\to 0^+}|I(t)|_{E(w)}
= 0$.  For any $\epsilon > 0$, choose $M$ so large such that
\begin{eqnarray}\label{E6.8}
\int^\infty_M e^ {-\eta} |v(\xi)| (1 + |\xi|^j)^{-1}d \eta \leq
\int^\infty_M e^ {-\eta }d\eta \cdot |v|_{E(w)} < {\epsilon \over
4},\\
\int^\infty_M e^ {-\eta} | v(\xi + \sqrt{4t\eta}) | (1 + |\xi +
\sqrt{4t\eta}|^j)^{-1} {1 + |\xi + \sqrt{4t\eta}|^j\over 1 + |\xi|^j}
d\eta\\
\leq C\int^\infty_Me^{-\eta} |v|_{E(w)} (1 + (\sqrt{4t\eta})^j)d\eta <
{\epsilon \over 4},\nonumber
\end{eqnarray}

For that fixed $M$, let $t_\epsilon$ be small such that for $0 < t <
t_\epsilon$,
\begin{equation}\label{E6.10}
\int^M_0 e^ {-\eta} | v(\xi + \sqrt{4t\eta})(1 + |\xi +
\sqrt{4t\eta}|^j)^{-1} - v (\xi) (1 + |\xi|^j)^{-1} | d\eta < {\epsilon
\over 4}.
\end{equation}
This is possible since $v(\xi) w^{-1}(\xi)$ is uniformly continuous.  Also
if
$t_\epsilon$ is even smaller, then
\begin{equation}\label{E6.11}
\begin{array}{l}
\int^M_0 e^{-\eta} | v (\xi + \sqrt{4t\eta})| \vert (1 + |\xi +
\sqrt{4t\eta} |^j)^{-1} - (1 + |\xi|^j)^{-1} | d\eta\\
\leq \int^M_0 e^{-\eta} |v|_{E(w)} (1 - {1 + |\xi + \sqrt{4t\eta}|^j\over 1
+
|\xi|^j} ) d \eta < {\epsilon \over 4},
\end{array}
\end{equation}
since ${|\xi + \alpha |^j - |\xi|^j\over 1 + |\xi|^j} \to 0$ uniformly with
respect to $\xi$ as $\alpha \to 0$.

{}From (\ref{E6.8}) - (\ref{E6.11}), we see $\displaystyle \lim _{t \to 0^+}
\displaystyle \sup _\xi I_1(\xi,t) = 0$.

\noindent (ii)  Let $X = E_{\Bbb R^+}(w), D_A = E^2_{\Bbb R^+}(w) \cap BC$,
and
$u_0 \in E^1_{\Bbb R^+}(w) \cap BC$.  Let $\tilde u_0$ be the even
extension of
$u_0$ to $\xi \in {\Bbb R}$.  Then $\tilde u_0 \in E_{\Bbb R}(w)$.  To find
$e^{At}u_0$, we solve
$$
\tilde u_t = u_{\xi\xi},\ \ \tilde u(0) = \tilde u_0.
$$
The restriction of $\tilde u(t)$ to $\xi \geq 0$ is $e^{At}u_0$.  Using the
results from (i), we find that $|t^{1/2} Ae^{At}u_0|_{E_{\Bbb R^+}(w)} \leq
|t^{1/2}D^2_\xi \tilde u(t)|_{E_{\Bbb R}(w)}$.  The latter approaches $0$
as $t
\to 0^+$.  Thus $u_0 \in D_A({1\over 2})$.

\noindent (iii) The other cases can be treated similarly.
\end{pf*} 

\bg{pf*}{Proof of \lemref{LB.7}}  (a) Let $\lambda \in {\Bbb C}$ be such that
$Re \lambda > - \sigma _0$.  Let $g \in X$ and consider $(L - \lambda ) u =
g$.  The equivalent system is
\begin{equation}\label{E6.12}
\begin{array}{lll}
u_\xi & = & v,\\
v_\xi & = & (\lambda I - M)u - Vv + g.
\end{array}
\end{equation}
Let $J = \left ( \begin{array}{ll}
0 & I\\
\lambda I-M & -VI\end{array} \right )$.  Since $Re\sigma(\lambda I - M) >
0$, it
can be verified that $J$ is hyperbolic with $n$-dimensional stable and
unstable
subspaces.  Assume that $\eta + i\omega \in \sigma (\lambda I - M)$, and $\mu
\in
\sigma J$.  Then $\mu = [-V \pm \sqrt{V^2+4\eta + 4i\omega}]/2$.  One can
verify
that
$Re\sqrt{V^2+4\eta +4i\omega}>Re\sqrt{V^2+4\eta}$. Thus, $|Re\mu|>\sqrt{V^2
+ 4\eta} -|V|$.  If $Re \lambda + \sigma _0 > \eta _0$ for some $\eta _0 > 0$,
then
let $\gamma _0 = \sqrt{V^2 + 4\eta _0} - |V|$, we have
$$
\begin{array}{lr}
|e^{J\xi}P_s| < Ke^{-\gamma_0\xi}, & \xi \geq 0,\\
|e^{J\xi}P_u| < Ke^{\gamma _0\xi}, & \xi \leq 0,
\end{array}
$$
for some $K > 0$, where $P_s$ and $P_u$ are the projections to stable and
unstable subspaces of ${\Bbb R}^{2n}$.  Therefore,  if $\gamma$ is the
constant used to define $w$, with $ | \gamma |< \gamma _0$,
\begin{equation}\label{E6.13}
\left ( \begin{array}{c} u \\ v \end{array} \right ) (\xi) =
\int^\xi_{-\infty}
e^{J(\xi-\eta)} P_s \left ( \begin{array}{c} 0 \\ g(\eta)\end{array} \right
)
d_\eta + \int^\xi _\infty e^{J(\xi-\eta)} P_u\left ( \begin{array}{c} 0 \\
g(\eta)\end{array} \right ) d\eta
\end{equation}
is the unique solution to (\ref{E6.12}) that is in $E^2_{\Bbb R}(w)$.  It
is
easy to show that $|u|_{E^2(w)} \leq C|g|_{E(w)}$.  Thus, $\lambda \in
\rho(L)$.
This proves the case $X = E_{\Bbb R}(w)$.

Now consider $X = E_{\Bbb R^\pm}(w)$ and $g \in X$.  Replace $g$ by its
even
extension $\tilde g$ and consider (\ref{E6.12}).  The solution $\left (
\begin{array}{c}
u\\ v\end{array} \right )$ can be expressed by (\ref{E6.13}) and is even in
$\xi \in {\Bbb R}$.  Consider its restriction on $\xi \in {\Bbb R}^\pm$,
then
$u \in E^2_{\Bbb R^\pm}(w) \cap BC$.

(b) From part (a), if $Re\lambda > - \sigma _0 + \eta _0$, equation
$u_{\xi\xi} + Vu_{\xi} + Df(p^i)u - \lambda u = 0$ has exponential
dichotomies
on ${\Bbb R}^-$ and ${\Bbb R}^+$.  Therefore
\begin{equation}\label{E6.13a}
\begin{array}{lll}
u_\xi & = & v\\
v_\xi & = & [\lambda I - Df(q)]u - Vv
\end{array}
\end{equation}
has exponential dichotomies on ${\Bbb R^-}$ and ${\Bbb R^+}$ with
projections
$P_s(t) + P_u(t) = I,\, t \in {\Bbb R^-}$ and ${\Bbb R^+}$ respectively.  If
$\lambda $ is such that ${\Bbb R}P_u(0^-)\cap {\Bbb R}P_s(0^+) = \{0\}$,
then
(\ref{E6.13a}) has exponential dichotomy on ${\Bbb R}$.  Similar to part (a),
we
can show $\lambda \in \rho(L_q)$.  If $\lambda$ is such that ${\cal R}P_u(0^-)
\cap {\cal R}P_s(0^+) \not= \{0\}$, then $\lambda $ is an eigenvalue of
finite multiplicity.
\end{pf*} 

\bg{pf*}{Proof of \lemref{L3.1}}     (i) The results  come from H1)
by the implicit function theorem and the continuous dependence of eigenvalues
on the parameter $x$.

(ii) Equation (\ref{E3.2}) has a heteroclinic solution $q(\xi)$
when $x
= x^i$ and $V^i(x^i)  = 0$.  Melnikov's method is used to determine the
existence
of heteroclinic solutions $(u,v)$ near $(q,q_\xi)$ for the following system
\begin{equation}\label{E6.14}
\begin{array}{lll}
u_\xi & = & v,\\
v_\xi & = & -f_0(u,x) - Vv.
\end{array}
\end{equation}
Here  $V \in {\Bbb R}$ and $x$ are parameters.  Based on H1), H2) and
\lemref{LB.1}, the linear
variational equations for (\ref{E6.14}) around $(u,v)=(q,q_\xi)$
has exponential dichotomies on
${\Bbb R^-}$ and ${\Bbb R^+}$ with ${\cal R}P_s(0^+) \cap {\cal R}P_u(0^-)$
spanned by $\{(q_\xi(0),q_{\xi\xi}(0))\}$, when $V = 0$ and $x = x^i$.
Let $G(x,V)$ denote the function
that measures the distance between $W^u(p^1)$ and $W^s(p^2)$
along the direction
$(\psi_{i1}(0),,\psi_{2i}(0))$ based on the point $(q(0), q_\xi(0))$.  cf
(\ref{E6.4a}).  Notice that $(\psi_{i1}(0),\psi_{2i}(0))$ is transverse to
$T_{q(0)} W^u(p^\prime) + T_{q(0)}W^s(p^2)$, thus, $G(x,V)$ is well
defined. From \cite{palmer84},
\begin{equation}\label{E6.14a}
\begin{array}{lll}
{\partial G(x^i,0)\over \partial x} & = &
\int^\infty_{-\infty} \psi_{i2}^\tau(\xi) f_{0x} (q^i(\xi),x^i)d\xi,\\
{\partial G(x^i,0)\over \partial V} & = & \int^\infty_{-\infty}
\psi^\tau_{i2}(\xi) q^i_\xi (\xi) d\xi.
\end{array}
\end{equation}
{}From (\ref{E3.1}), ${\partial G(x^i,0)\over \partial V} \not= 0$.
Therefore
there exist $V^i(x)$ such that $G(x,V^i(x)) = 0$ for $x$ in a neighborhood
of
$x^i$.  The method in \cite{palmer84} also insures that the heteroclinic
solution $q^i(\xi,x)$ depends smoothly on $x$.

(iii)  Define $L^i_x \varphi$ by the left hand side of (\ref{E3.3}). When
$x=x^i$, $L^i_x: C^2_{bu} \to C^0_{bu}$ is Fredholm with index equal zero, see
\cite{palmer84}. From \cite[page 115]{schechter}, if x is in a small
neighborhood of
$x^i$, $L_x^i$ is Fredholm with index zero. Also $\dim{\cal K}L^i_x \leq
\dim{\cal K}L^i_{x^i}$. Since $q_\xi \in {\cal K}L^i_x$, we have $\dim{\cal
K}L^i_x=1$ for all $x\in O^i$. Therefore (\ref{E3.4})
has a unique bounded solution $\psi_i(\xi,x)$ up to constant multiples.

It remains to show that by choosing $|\psi_i(\xi,x)|=1$, $\psi_i$ is a
smooth function of $x$.
Let $\{U_1 = (q_\xi(0),q_{\xi\xi}(0)), U_2, \ldots , U_n\}$ be an
orthogonal basis for ${\cal R}P_u(0^-)$ and $\{\overline U_1 = U_1, \overline
U_2, \ldots , \overline U_n\}$ be an orthogonal basis for ${\cal R}P_s(0^+)$
of the system
\begin{equation}\label{E6.15}
\begin{array}{lll}
u_\xi & = & v,\\
v_\xi & = & f_{0u}(q^i(\xi,x), x)u - V^i(x)v,
\end{array}
\end{equation}
when $x = x^i$ and $V^i(x) = 0$. System (\ref{E6.15}) has exponential
dichotomies on ${\Bbb R}^-$ and ${\Bbb R}^+$ when $x\in O^i$. Let
$P_u(x,t)$ and $P_s(x,t)$ denote the projections to the unstable and
stable spaces. $P_u(x^i,t)=P_u(t)$ and $P_s(x^i,t)=P_s(t)$.

Assume that $x$ is near $x^i$ so that $V^i(x)$ is near zero.  For each
$U_i,\, 2 \leq i \leq n$, there exists a unique $\Delta U_i \in
{\cal R}P_s(0^-)$ such
that $U_i + \Delta U_i \in {\cal R}P_u(x,0^-)$.  Also for each $\overline
U_i, 2 \leq i \leq n$, there exists a unique $\Delta \overline U_i \in
{\cal
R}P_u(0^+)$ such that $\overline U_i + \Delta \overline U_i \in {\cal
R}P_s(x,0^+)$. The functions $\Delta U_i$ and $\Delta \overline
 U_i$ are smooth functions of $x$. See \cite{halelin86} for details. For
$i=1$ let $U_1+\Delta U_1 = (q^i_\xi(0,x) , q^i_{\xi\xi}(0,x))$. $\Delta
U_1$ is also smooth in $x$.
In particular,  $|\Delta U_i| + |\Delta \overline U_i| = O(|x - x^i|)$ for
all $i$.  This proves that if $x-x^i$ is small,
$$
{\cal R}P_u(x,0^-) + {\cal R}P_s(x,0^+) = \text{span}(U_1 + \Delta
U_1\,\ldots U_n + \Delta U_n\,\overline U_2 +\Delta \overline U_2 \ldots
\overline U_n + \Delta \overline U_n),
$$
depends smoothly on $x$. The adjoint equation of (\ref{E6.15}) can
be found in (\ref{E6.4a}) where $V=V^i,\: q=q^i$. Let $(\psi_{i1}(x),
\psi_{i2}(x))$ be the unique bounded solution to the adjoint equation with
$|\psi_{i2}(0)|=1$. Let $U_0 = (\psi_{i1}(0), \psi_{i2}(0))$ that is
orthogonal to ${\cal R}P_u(0^-) + {\cal R}P_s(0^+)$.
By a standard projection method, there exists a unique
$\Delta U_0 \in {\cal R}P_u(x,0^-) + {\cal R}P_s(x,0^+)$ such that $U_0
+ \Delta U_0$ is orthogonal to ${\cal R}P_u(x,0^-) + {\cal
R}P_s(x,0^+)$. It can be shown that (\ref{E6.4a}) has a bounded solution with
the initial data $U_0+ \Delta U_0$, cf. \cite{halelin86}.
A normalized solution $(\psi_{i1}(0,x), \psi_{i2}(0,x))$ with the initial data
$(\psi_{i1}(0,x), \psi_{i2}(0,x)),\: |\psi_{i2}(0,x)|
=1$, can obtained by rescaling.  Let $\psi_{i}(0,x)= \psi_{i2}(0,x)$.
This proves the smooth dependence of $\psi_{i}(0,x)$ on $x$.

(iv) From Lemma B.7, (b), $\sigma \{L^i_x\} \cap \{Re \lambda > -
\sigma _0 + \eta _0\}$ consists of isolated
eigenvalues of finite order.  When $x = x^i,\, V^i(x) = 0$, from H7),$\lambda
= 0$ is a simple eigenvalue, all the other eigenvalues satisfy $Re \lambda <
- \alpha _0$.  Since eigenvalues depend continuously on $x$.  Thus, if $0 <
\overline \alpha  _0 = \min \{\alpha _0, \sigma _0 - \eta _0\}$, there
exists
$\epsilon > 0$ such that if $|x - x^i| < \epsilon$ then $\lambda = 0$ is
the
only eigenvalue in $\{Re \lambda  > - \overline \alpha _0\}$ and is simple.

(v)  From H6), (\ref{E3.5}) is valid if $x = x^i$.  Since both
integrals in (\ref{E3.5}) depend continuously on $x$, thus (\ref{E3.5}) is
valid for $|x - x^i| < \epsilon$ if $\epsilon > 0$ is small.  The formula
for ${\partial V^i(x^i)\over \partial x}$ follows from (\ref{E6.14a}).
\end{pf*} 

\bg{pf*}{Proof of \lemref{L3.2}}  $L^i_x$ is a Fredholm operator in
$E^m_{\Bbb R}(w)$ with index zero.  From \lemref{L3.1}, (iii), $Ker\{L^i_x\} =$
span$\{q^i_\xi(\cdot,x)\}$ is one-dimensional.  Therefore Range$\{L^i_x\}
\newline =
\{\psi_i(\cdot,x)\}$ is of codimension one.  Consider the mapping ${\cal
F}:(x,u,V_1) \to (g,h)$ as follows
\begin{equation}\label{E6.17}
\begin{array}{l}
L^i_xu - V_1q^i_\xi(\cdot , x) = g,\\
<u,q^i_\xi(\cdot,x)> = h.
\end{array}
\end{equation}
${\cal F}:O^i \times E^{m+2}_{\Bbb R}(w) \times {\Bbb R} \to E^m_{\Bbb
R}(w)
\times {\Bbb R}$ is $C^\infty$, in fact, linear with respect to $u \in
E^{m+2}_{\Bbb R}(w)$ and $V_1 \in {\Bbb R}$.  It can be verified that
$\partial  {\cal F}/\partial (u,V_1)$ is a linear homeomorphism in the
indicated norms.  We only need to show that $(u,V_1)$ is uniquely solvable
form (\ref{E6.17})
for any $(g,h) \in  E^m_{\Bbb R}(w) \times {\Bbb R}$.  If we
choose $V_1 =
<\psi_i(\cdot , x),g > \cdot < \psi_i(\cdot,x),q^i_\xi(\cdot,x)>^{-1}$,
then $
g - V_1  q^i_\xi \in$ Range $\{L^i_x\}$.  Any two solutions of the first of
(\ref{E6.17}) differ by a multiple of $q^i_\xi \in Ker\{L^i_x\}$, that can
be
determined by the second of (\ref{E6.17}).  Let $h = 0$ and denote the
solutions by $V_1 = V^i_*(x,g)$ and $u = u(\cdot,x,g)$.

The smoothness of $V^i_*(x,g)$ and $u(\cdot,x,g)$ on $(x,g)$ also
follows from the Implicit Function Theorem applied on the function
${\cal F}$.
\end{pf*} 

\bg{pf*}{Proof of \lemref{L4.1}}  According to \lemref{LB.7}, (b), ${\Bbb C}_1
=  \sigma \{L_{q^i}\} \cap \{Re \lambda > - \partial _0 + \eta _0\}$
consists
of only eigenvalues.  When $i = 0, r$, from H5), $L_{q^i}$ has no
eigenvalues
in ${\Bbb C}_1$ in the space $C_{bu}({\Bbb R}^+,{\Bbb R}^n)$.  Thus, it
also
has no eigenvalues in ${\Bbb C}_1$ in the space $B_{{\Bbb R}^\pm}(w)$.  This
proves the case $i = 0,r$.  When $1 \leq i \leq r -1$, from H5) again,
in
$C_{bu}({\Bbb R},{\Bbb R}^n)$, the only eigenvalue of $L_{q^i}$ in ${\Bbb
C}_1$
is $\lambda = 0$, simple.  Thus the only eigenvalue of $L_{q^i}$ in $B_{\Bbb
R}(w)$ is also $\lambda = 0$, simple.  From \lemref{LB.6}, $q^i(\cdot,\overline
\eta ^i)$ is asymptotically stable modulo spatial shifts.
\end{pf*} 

\bg{pf*}{Proof of \lemref{L4.2}}  The proof is exactly like that of
\lemref{L4.1}.
\end{pf*}

\endinput


\makeatletter \renewcommand{\@biblabel}[1]{\hfill#1.}\makeatother

\endinput